\documentclass[fleqn,usenatbib,useAMS]{mnras}

\usepackage{graphicx}
\usepackage{bm}
\usepackage{amsmath}	
\usepackage{amssymb}	
\usepackage{multicol}        
\usepackage{bm}		
\usepackage{pdflscape}	
\usepackage{subcaption}
\usepackage{tabularx}
\usepackage{ulem}

\usepackage[T1]{fontenc}
\usepackage{ae,aecompl}
\usepackage{newtxtext,newtxmath}

\title[Reconstructed velocity comparison with SN]{Cosmic flows in the nearby Universe: new peculiar velocities from SNe and cosmological constraints}

\author[S. S. Boruah, M. J. Hudson and G. Lavaux]{Supranta S. Boruah$^{1, 2}$\thanks{Contact e-mail: \href{mailto:ssarmabo@uwaterloo.ca}{ssarmabo@uwaterloo.ca}}, Michael J. Hudson$^{1,3,4}$\thanks{Contact e-mail: \href{mailto:mike.hudson@uwaterloo.ca}{mike.hudson@uwaterloo.ca}}, Guilhem Lavaux$^{5}$\thanks{Contact e-mail: \href{mailto:guilhem.lavaux@iap.fr}{guilhem.lavaux@iap.fr}}
\\
$^{1}$Waterloo Centre for Astrophysics, University of Waterloo,  200, University Ave W, Waterloo, ON N2L 3G1\\
$^{2}$Department of Applied Mathematics, University of Waterloo,  200, University Ave W, Waterloo, ON N2L 3G1 \\
$^{3}$Department of Physics and Astronomy, University of Waterloo, Waterloo, ON, N2L 3G1, Canada\\
$^{4}$Perimeter Institute for Theoretical Physics, 31 Caroline St N, Waterloo, ON N2L 2Y5\\
$^{5}$CNRS \& Sorbonne Universit\'e, UMR7095, Institut d'Astrophysique de Paris, F-75014, Paris, France 
}
\date{Last updated XXXX; in original form YYYY}

\pubyear{2019}

\begin{document}
\label{firstpage}
\pagerange{\pageref{firstpage}--\pageref{lastpage}}
\maketitle

\newcommand{\diffop}{\mathrm{d}}
\newcommand{\borg}{{\sc borg}}
\newcommand{\gl}[2]{\textcolor{red}{[GL: \sout{#1} #2]}}
\newcommand{\mmat}[1]{{\mathbf{#1}}}
\newcommand{\mvec}[1]{{\bm{#1}}}
\newcommand{\fsigmainv}{0.400 \pm 0.017}
\newcommand{\seightinv}{0.776\pm0.033}
\newcommand{\vextinv}{171^{+11}_{-11}}
\newcommand{\lextinv}{301^{\circ} \pm 4^{\circ}}
\newcommand{\bextinv}{0^{\circ} \pm 3^{\circ}}
\newcommand{\vbulkinv}{252 \pm 11}
\newcommand{\lbulkinv}{293^{\circ} \pm 5^{\circ}}
\newcommand{\bbulkinv}{14^{\circ} \pm 5^{\circ}}
\begin{abstract}
The peculiar velocity field offers a unique way to probe dark matter density field on large scales at low redshifts. In this work, we have compiled a new sample of 465 peculiar velocities from low redshift $(z < 0.067)$ Type Ia supernovae. We compare the reconstructed velocity field derived from the 2M++ galaxy redshift compilation to the supernovae, the SFI++ and the 2MTF Tully-Fisher distance catalogues. We used a forward method to jointly infer the distances and the velocities of distance indicators by comparing the observations to the reconstruction. Comparison of the reconstructed peculiar velocity fields to observations allows us to infer the cosmological parameter combination $f\sigma_8$, and the bulk flow velocity arising from outside the survey volume. The residual bulk flow arising from outside the 2M++ volume is inferred to be $\vextinv$ km s$^{-1}$ in the direction $l=\lextinv$ and $b=\bextinv$. We obtain $f\sigma_{8} = \fsigmainv$, equivalent to $S_8 \approx \sigma_8(\Omega_m/0.3)^{0.55}=\seightinv$, which corresponds to an approximately $ 4\%\,$ statistical uncertainty on the value of $f\sigma_8$. Our inferred value is consistent with other low redshift results in the literature.
\end{abstract}

\begin{keywords}
Galaxy: kinematics and dynamics -- galaxies: statistics -- large-scale structure of Universe -- cosmology: observations
\end{keywords}



\section{\label{sec:intro} Introduction}

Peculiar velocities, the deviation from the regular Hubble flow of the galaxies, are sourced by inhomogeneities in the universe, making them an excellent probe of its large-scale structure. In fact, the peculiar velocity field is the only probe of very large-scale structures in the low-redshift universe. 

In linear perturbation theory, the relationship between peculiar velocity, $\mvec{v}$, and the dark matter overdensity, $\delta$, is given as
\begin{equation}
	\mvec{v}(\mvec{r}) = \frac{H_0 f}{4\pi} \int \diffop^3 \mvec{r}^{\prime} \delta(\mvec{r}^{\prime}) \frac{\mvec{r}^{\prime} - \mvec{r}}{|\mvec{r}^{\prime} - \mvec{r}|^3}\,.
	\label{eqn:linear}
\end{equation}
where $\delta = \rho/\bar{\rho} - 1$, with $\rho$ being the density and $\bar{\rho}$ the mean density of the Universe. As can be seen from the above equation, the velocity field is sensitive to the dimensionless growth rate, $f = \frac{\diffop \ln D}{\diffop\ln a}$ and the typical amplitude of density fluctuations. Here, $D$ is the growth function of linear perturbations, and $a$ is the scale factor. Consequently, the peculiar velocity field has been used to constrain the degenerate cosmological parameter $f \sigma_8$ \citep{pike_hudson, Davis11, Carrick_et_al, 6dF_cross_correlation, cosmicflows_growth_rate}, where $\sigma_8$ is the root mean squared fluctuation in the matter overdensity in a sphere of radius $8~h^{-1}$ Mpc. In the $\Lambda$CDM cosmological model, $f \approx \Omega^{0.55}_m$ \citep{wang_steinhardt}. However, in modified theories of gravity, the growth rate could be different, i.e. $f = \Omega^{\gamma}_m$ with $\gamma \neq 0.55$. Therefore, peculiar velocities can also be used to constrain theories of gravity \citep{lcdm_test_lahav, nusser_lcdm, hudson_turnbull, Huterer17}.

However, analysing peculiar velocities poses several challenges. The measured redshift, $cz$, of a galaxy gets a contribution from both the recessional velocity due to Hubble flow, $Hr$, and the peculiar velocity, $\mvec{v}$. Therefore, to analyse peculiar velocities, one needs to determine the distances to these galaxies in order to separate these two contributions. There are several ways to measure distances directly. The most popular of these use empirical galaxy scaling relations. For example, SFI++ \citep{sfi1} and the 2MTF \citep{2MTF} catalogues use the Tully-Fisher (TF) relation \citep{TF_relations}, and the 6dF velocity survey \citep{6dfv} uses the Fundamental Plane (hereafter FP) relation \citep{FP_relation1, FP_relation2}. Another distance indicator relies on Type Ia supernovae \citep{supernovae_pec_vel1, supernovae_pec_vel2, turnbull_sn, Huterer17, sn1a_vel3}. Since the distance errors from Type Ia supernovae ($\mathcal{O}[5$-$10 \%]$) are much lower than those obtained using galaxy scaling relationships ($\mathcal{O}[20$-$25 \%]$), a smaller sample of Type Ia supernovae can give comparable results to that of a larger catalogue based on the TF or FP relations. In this work, we combine low redshift supernovae from various surveys to produce the largest peculiar velocity catalogue based on Type Ia supernovae to date.

The different approaches to analysing peculiar velocities can be separated into two categories: i) those which use only the distance indicator data for peculiar velocity analysis, and ii) those which `reconstruct' the cosmic density field from a redshift survey and then compare the velocity field predictions with the observed peculiar velocity data. Some examples of the first category are the POTENT \citep{potent} and the forward-modelled VIRBIUS \citep{virbius} method. Our approach falls into the second category, where we compare the reconstructed velocity field to distance observations. In particular, we use the distribution of galaxies $(\delta^g)$ as a tracer of the mass density field, $\delta$. We can then use a modified version of Equation~\eqref{eqn:linear} to predict the peculiar velocities. In this approach, we can constrain the degenerate parameter combination $\beta = f/b$, where $b$ is the linear galaxy bias. The cosmological parameter combination $f\sigma_8$ is then related to $\beta$ as $f\sigma_8 = \beta \sigma^g_8$, where, $\sigma^g_8$ is the typical fluctuation in the galaxy overdensity field at a radius of $8~h^{-1}$ Mpc. 
Specifically, we compare the observed data from the peculiar velocity surveys to the reconstructed velocity field from the 2M++ redshift compilation. In doing so we use an inverse reconstruction scheme which was used in \cite{Carrick_et_al}. More examples of reconstruction-based peculiar velocity analyses are given in \citet{mak} and \citet{2mass_reconstruction}.

This paper is structured as follows: Section~\ref{sec:pec_data} describes the peculiar velocity catalogue we use in this work, primarily, the new compilation of type Ia supernovae. Section~\ref{sec:reconstruction} describes the 2M++ galaxy catalogue and the reconstruction scheme used in this work. In Section~\ref{sec:compare}, we elaborate on the methods used to compare the reconstructed velocity field to the observations of the peculiar velocity catalogue. The results are presented in section \ref{sec:results}. We compare our results to other results in the literature and discuss future prospects of peculiar velocity analysis in section \ref{sec:discussion} before we summarise our results in section \ref{sec:summary}. We show that the modified forward likelihood method gives unbiased results on simulations in Appendix \ref{sec:sn_mock}. Throughout this work, $h = H_0/(100$ km s$^{-1}$ Mpc$^{-1}$), where $H_0$ is the local Hubble constant. Unless otherwise mentioned, throughout this work, we adopt a flat $\Lambda$CDM cosmological model with $\Omega_m = 0.3$.

\section{Peculiar velocity catalogue}\label{sec:pec_data}

In this section, we describe the two main peculiar velocity catalogues used in this paper. In Section~\ref{ssec:supernova_data}, we present a new compilation of low redshift Type Ia supernovae from various different surveys. In Section~\ref{ssec:tf}, we summarise the data from the SFI++ and 2MTF catalogues, where the distance has been estimated using the Tully-Fisher relationship.

\begin{figure}
	\includegraphics[width=\linewidth]{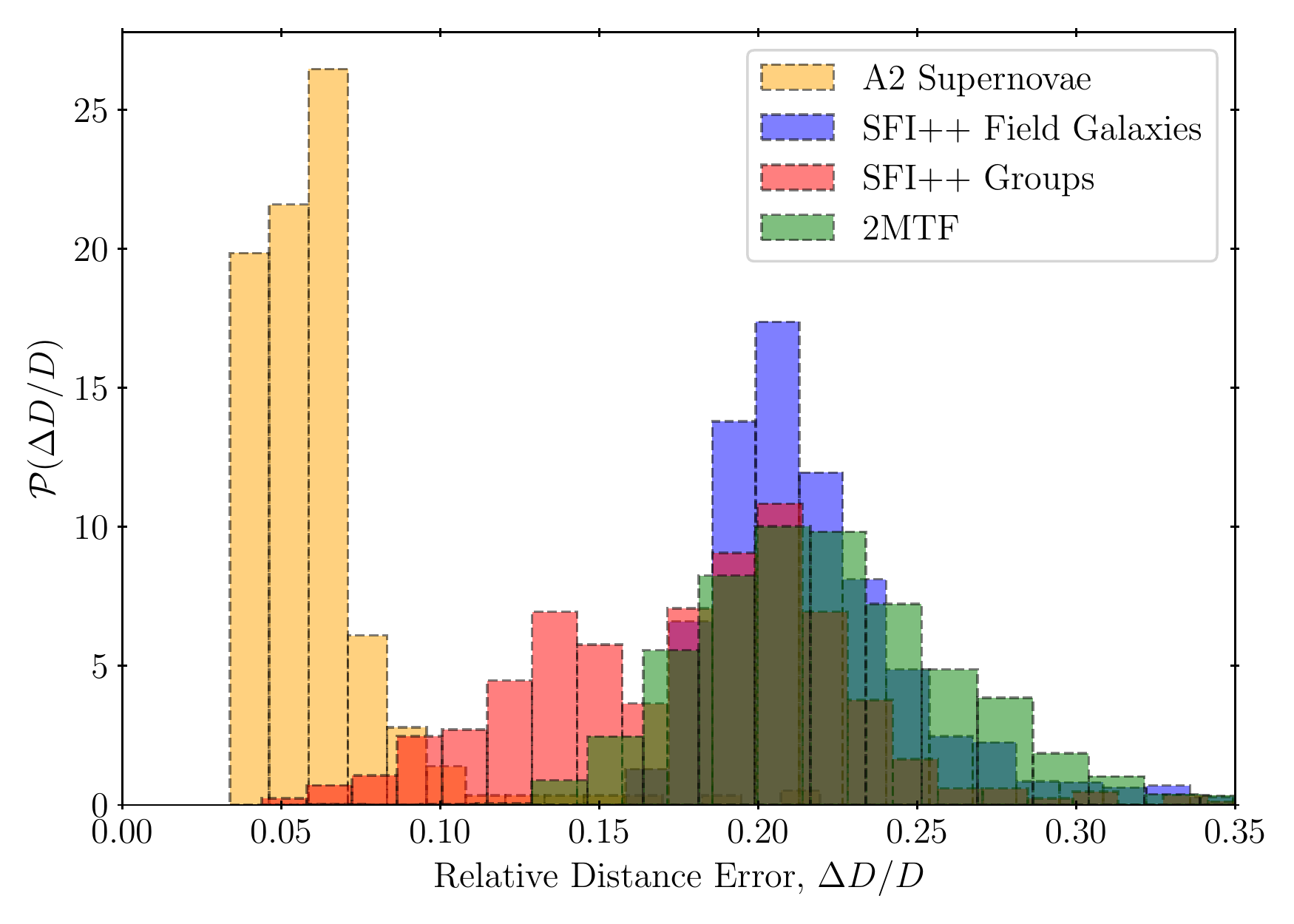}
	\caption{The normalised distribution of the relative distance errors of the different peculiar velocity datasets. The distribution of the relative errors of the A2 compilation is shown in orange, that of the SFI++ groups is shown in red, the SFI++ field galaxies are shown in blue and the 2MTF galaxies are shown in green. The typical errors on the distances of the supernovae is much lower than that of the other datasets.}
	\label{fig:dist_errors}
\end{figure}

\subsection{Second Amendment (A2) supernovae compilation}\label{ssec:supernova_data}

 Several distance indicators have been used over the past decades. Among these, distances from the FP and TF relations have found a central place in peculiar velocity analysis. In this section, we focus on distances derived from SNe-Ia light curves. Even with the recent addition of gravitational wave ``standard siren'' distances \citep[see e.g.][]{gw_distance}, SNe distances (typically $\sim 5\%$) are still the best in terms of distance errors (see Figure \ref{fig:dist_errors}). The peak luminosity of a Type Ia supernova is correlated with the rate of decline of its light curve, making these `standardisable candles' \citep{sn_standardisable_candle}. Type Ia supernovae have been previously used in many works to probe the velocity field of the nearby universe \citep{supernovae_pec_vel1, supernovae_pec_vel2, turnbull_sn, Huterer17}. 

\cite{turnbull_sn} previously presented the `First Amendment' (A1) compilation of Type Ia supernovae in the local universe. This was based on the addition of 26 SNe from the first data release (DR1) of the Carnegie Supernovae Project  \citep[CSP,][]{CSP_sn1_folatelli} to the low-$z$ set of `Constitution' supernovae \citep{constitution}. In this work, we add to the First amendment catalogue additional supernovae from the third data release (DR3) of CSP \citep{CSP_DR3}, the Lick Observatory Supernova Search \citep[LOSS,][]{LOSS} and the Foundation Supernova Survey data release 1 (DR1) \citep{foundation1, foundation2}, resulting in the `Second Amendment' (A2) compilation of SNe peculiar velocities. For each of these sub-catalogues, we only use the supernovae that are within the 2M++ volume and remove the duplicates from different catalogues. We also do a simple $\chi^2$ fit (described in section \ref{ssec:chisquared}) to determine the `flow model', which refers to the set of four parameters consisting of the rescaling factor, $\beta$  and the three components of the residual bulk flow velocities. We then reject iteratively the outliers from this fit until there are no outliers. While it is common to choose a fixed rejection threshold for all samples (such as $3\sigma$), in principle the threshold for outlier rejection should depend on the size of the sample. Suppose that the residuals are Gaussian then we expect a given object to lie in the tail of the distribution with some small probability $p$. If there are $N$ objects in the sample, the number of objects in the tails will be $Np$. We choose the threshold $p$ such that $Np = 0.5$. For the different samples, this value is  $2.6\sigma$ for the LOSS and CSP-DR3 samples and $2.9\sigma$ for the Foundation sample.  The LOSS supernova sample was taken from \citet{LOSS_data}. Removing the duplicates from the A1 catalogue and outliers from the $\chi^2$ fit, we arrive at a sample of $55$ SNe. We reject a total of $4$ outliers: (SN2005ls, SN2005mc, SN2006on, SN2001e) in the process. The data set for the CSP-DR3 was obtained from \citet{CSP_DR3_data}. From this catalogue, we remove the duplicates in the First Amendment or the LOSS sample. We also remove supernovae outside the 2M++ volume or the outliers of the $\chi^2$ fit. This yields us a total of $53$ supernovae after rejection of $2$ outliers (SN2006os, SN2008gp). Note that changing the duplicate selection criteria among A1, LOSS and CSP-DR3 changes the inferred value of $\beta$ by less than $0.1\sigma$. We use the given selection criteria for easy comparison with earlier results \citep{Carrick_et_al, turnbull_sn}. For the Foundation DR1 sample we rejected a total of $12$ outliers in the fitting procedure: (SN2016cck, SN2016gkt, ASASSN-15go, ASASSN-15mi, PS15akf, SN2016eqb, SN2016gfr, SN2017cju, ASASSN-15la,PS15bbn, SN2016aqs, SN2016cyt).

The Foundation DR1 and the LOSS sample provides the supernovae light-curve stretch parameter, $x_1$, color parameter $c$ and the amplitude, $m_B$ after fitting the light curves using the SALT2 \citep{SALT2} fitter. The distance modulus for the SALT2 model is given by the Tripp formula \citep{tripp},
\begin{equation}\label{eqn:tripp}
    \mu = m_B - M + \alpha x_1 - \mathcal{B} c \,.
\end{equation}
To determine the global parameters, we use a self consistent method to jointly fit for the flow model and the global parameters to determine the distances. We fit for the global parameters $\alpha, \mathcal{B}$, $M$ and the intrinsic scatter for this sample of supernovae in addition to the flow model using a modified forward likelihood method, which is described in section \ref{sssec:fwd_lkl_joint}. We note that, conventionally, the parameter $\mathcal{B}$ is denoted with $\beta$ in the SN literature. We avoid this notation to avoid confusion with $\beta = f/b$. 

It is useful to have a metric that summarise the power of a sample. Here we discuss two such metrics. First, we define a sample's uncertainty-weighted ``characteristic depth'' as
\begin{equation}
    d_{*} = \frac{\sum_{i=1}^{N} r_i / \sigma^2_i }{\sum_{i=1}^{N} 1 / \sigma^2_i }\,,
\end{equation}
where $\sigma_i$ is the uncertainty in the distance estimates. The characteristic depth of the different sub-samples of supernovae and the combined catalogue is presented in Table~\ref{tbl:sn_survey_properties}. Note that the newly added LOSS and Foundation samples probe higher redshifts compared to the earlier A1 sample. The characteristic depth of the full A2 sample is $41\ h^{-1}$ Mpc.
Second, we introduce a metric designed to assess the overall power of a peculiar velocity survey that combines the number, depth and uncertainty of the peculiar velocity tracers. If all galaxies in the sample were located in the same direction in the sky, and there was a bulk flow in that direction, then the uncertainty in the mean peculiar velocity, $\Delta \bar{V}$, of the sample is simply 
\begin{equation}
    \frac{1}{\Delta \bar{V}^2} = \sum_{i=1}^{N} \frac{1}{\sigma^2_i + \sigma^2_v},
\end{equation}
where the distance uncertainty, $\sigma_i$, is reported in km s$^{-1}$ and $\sigma_v$ quantifies the uncertainty in the peculiar velocity arising from non-linear perturbations. We assume a value of, $\sigma_v = 150$ km s$^{-1}$ which was obtained in \citet{Carrick_et_al} by comparing to $N$-body simulations. $\Delta \bar{V}$, is related to the expected error on bulk flow that can be obtained using a given peculiar velocity sample and hence is a Figure-of-merit for the peculiar velocity sample. We show the value of $\Delta \bar{V}$ for the different samples in Table \ref{tbl:sn_survey_properties}.

\begin{figure}
    \centering
    \includegraphics[width=\linewidth]{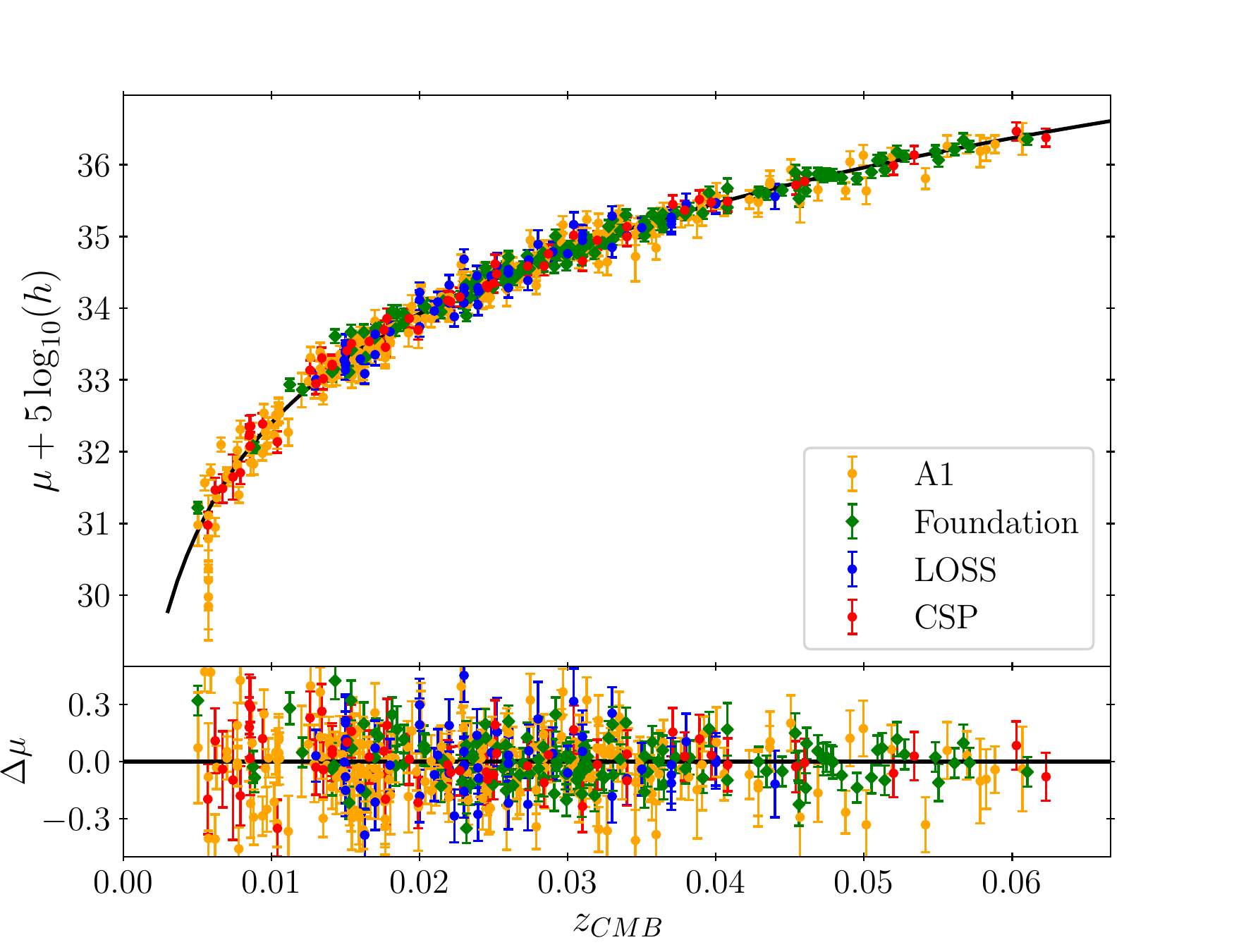}
    \caption{The Hubble diagram for the supernovae in the A2 compilation. The error bars for the magnitude include the intrinsic scatter for each sample. The black solid line is the expected distance-redshift relation in a $\Lambda$CDM cosmological model with $\Omega_m = 0.30$. The lower panel shows the residual from the given relation.}
    \label{fig:sn_hubble}
\end{figure}

\begin{figure}
    \centering
    \includegraphics[width=\linewidth]{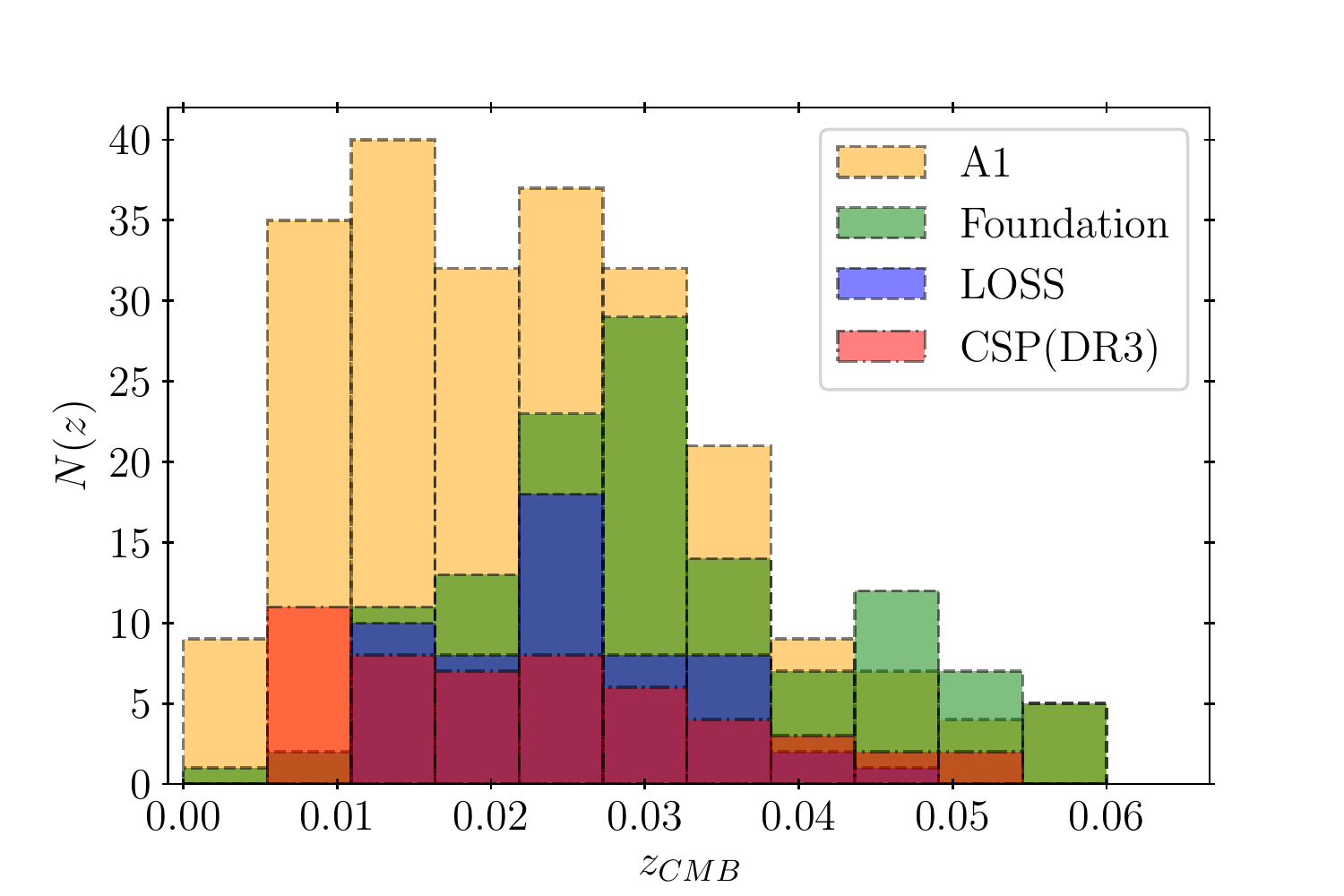}
    \caption{The redshift distribution of the supernovae in the different catalogues in the A2 compilation. Note that the LOSS and the Foundation samples probe higher redshifts, i.e., they have a higher characteristic depth ($r_{*} \sim 60~h^{-1}$ Mpc) compared to the A1 and CSP samples. The characteristic depth is shown in Table \ref{tbl:sn_survey_properties}.}
    \label{fig:sn_redshift}
\end{figure}

\begin{figure}
    \centering
    \includegraphics[width=\linewidth]{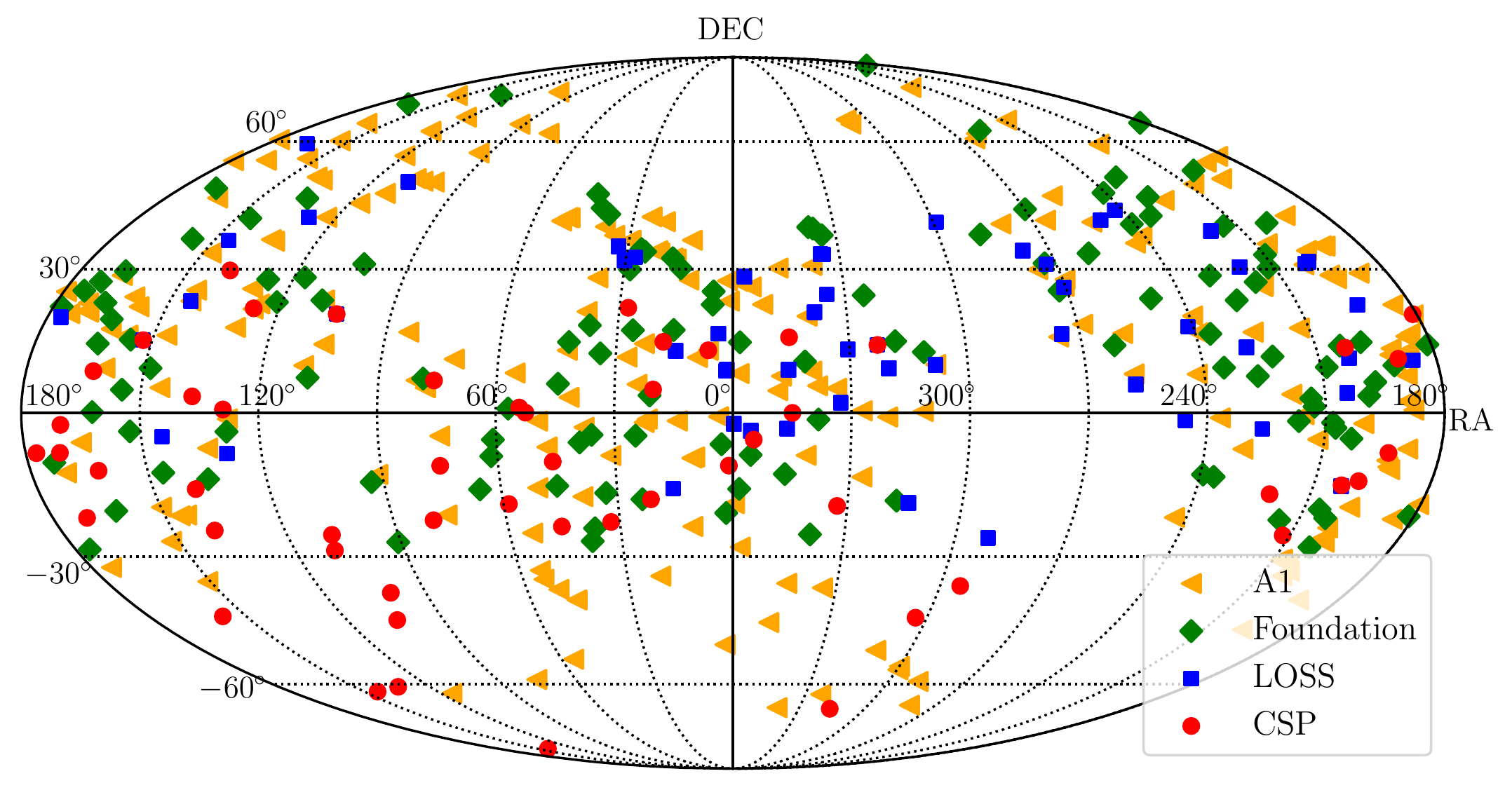}
    \caption{The sky distribution of the A2 supernovae in Equatorial coordinates. The above shows the Mollweide projection of the right ascension and the declination of the supernovae in the different samples. As can be seen in the figure, the CSP sample is primarily in the southern hemisphere.}
    \label{fig:sn_sky}
\end{figure}

The second amendment (A2) catalogue is included with the online supplementary material. A shortened version of the catalogue is shown in Table \ref{tbl:A2_cat}. A Hubble diagram for the supernovae in our compilation is shown in Figure \ref{fig:sn_hubble}. The redshift distribution of the supernovae in the A2 compilation is shown in Figure \ref{fig:sn_redshift}. The distribution of the supernovae on the sky is shown in a Mollweide projection in Figure \ref{fig:sn_sky}. Note that the CSP sample is distributed primarily in the southern sky.

\begin{table}
	\centering
	\caption{Properties of the different peculiar velocity catalogues showing the number of objects, the characteristic depth and the uncertainty in the mean peculiar velocity.}
	\begin{tabular}{lccc} 
		\hline
		Catalogue & $N_{\text{objects}}$ & $d_{*}~(h^{-1}$ Mpc) & $\Delta \bar{V}$ (km/s)\\
		\hline
		A1 & 232 & 31 & $28$\\
		CSP (DR3) & 53 & 40 & $49$\\
		LOSS & 55 & 61 & $85$\\
		Foundation & 125 & 59 & $38$\\

		\textbf{A2} & \textbf{465} & \textbf{41} & \textbf{16}\\

		SFI++ field galaxies & $1996$ & $40$ & $24$ \\
		SFI++ groups & $599$ & $22$ & $21$ \\
		2MTF & $1247$ & $26$ & $21$ \\
		\textbf{Combined} & $\boldsymbol{4308}$& $\boldsymbol{27}$ & $\boldsymbol{13}$\\
		\hline
	\end{tabular}
	\label{tbl:sn_survey_properties}
\end{table}

\begin{table*}
\caption{\label{tbl:A2_cat}Second Amendment compilation. Here we show a shortened version of the catalogue. The full table is available online.}
\begin{center}
\begin{tabular}{lrrrrrrrrrrrlr}
\hline
	ID	&$z_{\text{CMB}}$	&$\mu + 5\log_{10} h$	&$\Delta \mu$	& RA ($^{\circ}$)	& DEC ($^{\circ}$)	& $m_B$	& $\Delta m_B$	& $x_1$	& $\Delta x_1$& $c$	& $\Delta c$ & Sample	& GID\\

\hline
2006sr	&$0.0232$	&$34.312$	&$0.125$	&$0.931$	&$23.168$	&---	&---	&---	&---	&---	&---	&A1	&$   0$\\
2008hj	&$0.0379$	&$35.397$	&$0.040$	&$1.008$	&$-10.831$	&---	&---	&---	&---	&---	&---	&CSP	&$   0$\\
2003it	&$0.0240$	&$34.314$	&$0.142$	&$1.480$	&$27.488$	&---	&---	&---	&---	&---	&---	&A1	&$   0$\\
2000dg	&$0.0370$	&$35.214$	&$0.151$	&$1.558$	&$8.888$	&$16.989$	&$0.068$	&$-1.442$	&$0.420$	&$-0.057$	&$0.045$	&LOSS	&$   0$\\
2002hw	&$0.0160$	&$33.290$	&$0.140$	&$1.704$	&$8.630$	&$16.646$	&$0.038$	&$-1.849$	&$0.203$	&$0.377$	&$0.034$	&LOSS	&$   0$\\
2016ick	&$0.0512$	&$36.085$	&$0.078$	&$1.763$	&$-20.697$	&$17.280$	&$0.040$	&$0.430$	&$0.210$	&$-0.072$	&$0.031$	&Fdn	&$   0$\\
1992au	&$0.0607$	&$36.344$	&$0.222$	&$2.626$	&$-50.023$	&---	&---	&---	&---	&---	&---	&A1	&$   0$\\
ASASSN-15sf	&$0.0244$	&$34.562$	&$0.078$	&$2.865$	&$-6.427$	&$15.700$	&$0.040$	&$1.010$	&$0.140$	&$-0.065$	&$0.030$	&Fdn	&$   0$\\
1998dk	&$0.0120$	&$32.845$	&$0.238$	&$3.621$	&$-0.781$	&---	&---	&---	&---	&---	&---	&A1	&$   0$\\
2007s1	&$0.0260$	&$34.534$	&$0.137$	&$3.749$	&$16.335$	&$16.501$	&$0.028$	&$-0.746$	&$0.090$	&$0.021$	&$0.027$	&LOSS	&$   0$\\
...\\
\hline
\end{tabular}
\end{center}
$z_{\text{CMB}}$ is the redshift in the CMB Frame, corrected to the mean group redshift if appropriate. $\mu$ is the distance modulus for $h=1$, and $\Delta \mu$ its uncertainty. $m_B, x_1$ and $c$ are the Tripp parameters for the supernova and their corresponding uncertainties. The supernova sample (A1/LOSS/CSP-DR3 or Foundation) is denoted with Sample. GID denotes Group Identifier from the 2M++ catalogue. If there is no identified group in the 2M++ group catalogue, it is denoted with `$0$'. 
\end{table*}
\subsection{Tully-Fisher catalogues}
\label{ssec:tf}

The \citet{TF_relations} relation is an empirical scaling relationship between the luminosity and the velocity width of spiral galaxies. It is commonly expressed in terms of the variable, $\eta = \log W - 2.5$, where $W$ is the velocity width of the galaxies in km s$^{-1}$. This relationship can be used to determine distances to galaxies. The distance modulus to a galaxy in terms of the apparent magnitude $(m)$, and $\eta$ is given as,
\begin{equation}\label{eqn:tf_distance}
    \mu = m - (a_{\text{TF}} + b_{\text{TF}} \eta)\,,
\end{equation}
where $a_{\text{TF}}$ and $b_{\text{TF}}$ are the zero-point and the slope of the Tully-Fisher relationship. The intrinsic scatter is denoted with $\sigma_{\text{int}}$. 

For the analysis of the Tully-Fisher samples, we jointly fit for the distances and the flow model using the method described in section \ref{sssec:fwd_lkl_joint}. This requires fitting for three additional TF parameters, $a_{\text{TF}}$, $b_{\text{TF}}$ and $\sigma_{\text{int}}$ in addition to the flow model. In this work, we use the data from two TF catalogues: the SFI++ catalogue and the 2MASS Tully-Fisher (2MTF) survey. We present the details of data processing for the two catalogues in the next subsections. The value of the TF parameters for the SFI++ and the 2MTF catalogues as inferred in our fitting procedures are presented in Table \ref{tbl:tf_parameters}.

\begin{figure}
    \centering
    \includegraphics[width=\linewidth]{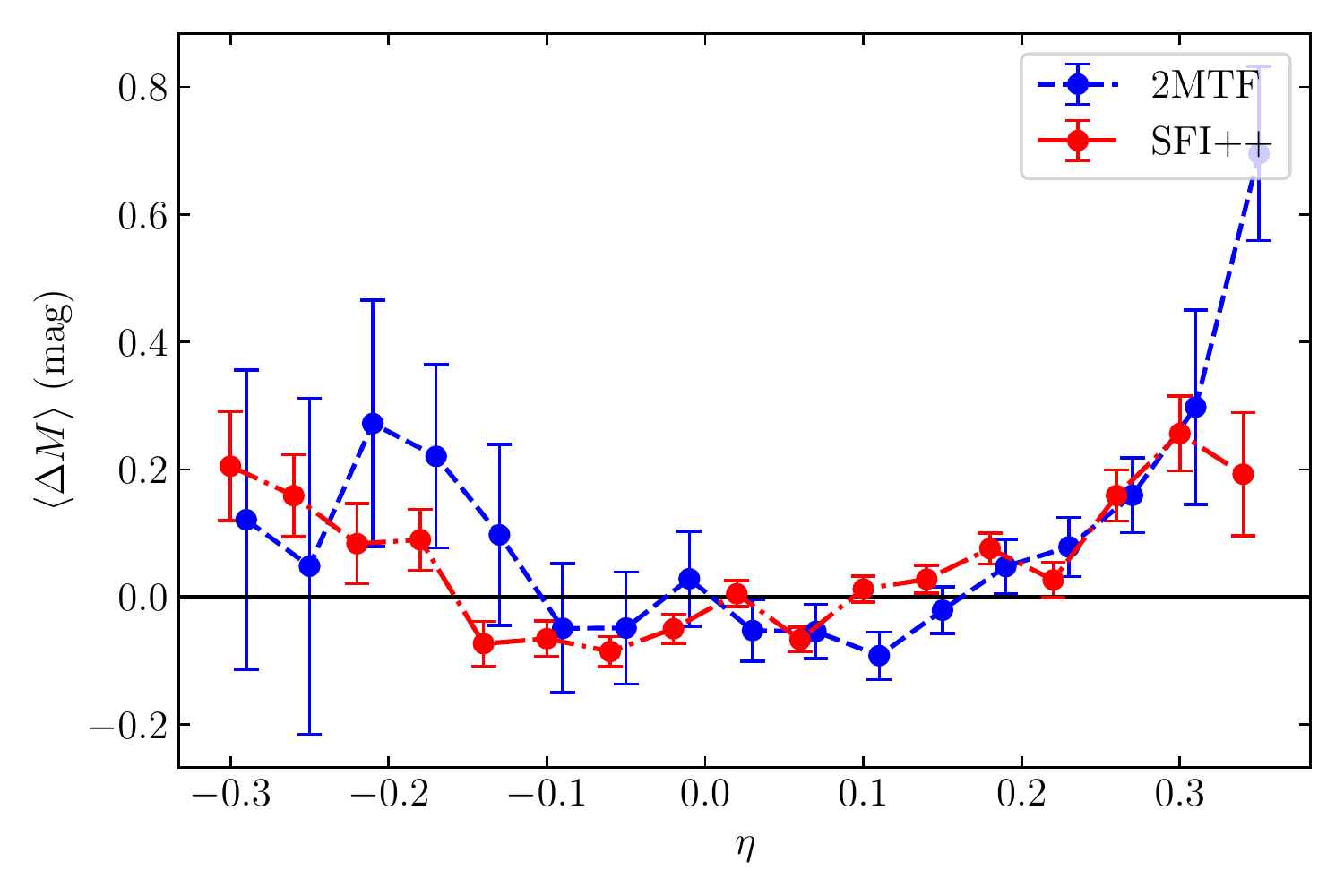}
    \caption{The deviation from the inferred linear Tully-Fisher relationship in bins of $\eta$. We calculated the mean absolute magnitude in bins of $\eta$ of width $0.04$ and calculate its deviation from the inferred linear relationship (shown on the $y$-axis). As can be seen, it deviates from the linear relationship in both in the faint end (low $\eta$) and in the bright end (high $\eta$).}
    \label{fig:tf_deviation}
\end{figure}

\begin{table}
	\centering
	\caption{Tully-Fisher parameters inferred using our fitting procedure}
	\label{tbl:tf_parameters}
	\begin{tabular}{lccc} 
		\hline
		 & $a_{\text{TF}} + 5\log_{10}(h)$ & $b_{\text{TF}}$ & $\sigma_{\text{int}}$ (mag)\\		
		\hline
		SFI++ & $-20.915\pm 0.008$ & $-6.42\pm0.07$ & $0.299 \pm 0.006$\\
		2MTF & $-22.556\pm0.013$ & $-6.56\pm0.13$ & $0.392 \pm 0.010$\\
	    \hline
	\end{tabular}
\end{table}
\subsubsection{SFI++}
The SFI++ catalogue \citep{sfi1, sfi2} consists of $4052$ galaxies and $736$ groups. After restricting to the groups and galaxies inside the region covered by 2M++, we are left with $3915$ galaxies and $734$ groups. For the set of galaxies, we then use the redshift distance as the distance estimate and fit for the Tully-Fisher relations. It was noted in \citet{Davis11} that the $I$-band Tully-Fisher relation deviates from a linear relationship at the faint end. Since we are fitting using a forward method, the selection cuts should be a function of $\eta$ only for an unbiased estimate. We plot the mean relation as inferred from the data and how it deviates from the inferred linear relationship in Figure \ref{fig:tf_deviation}. As can be seen from the figure, there is a deviation from the inferred linear relationship in both the faint end (low $\eta$) and the bright end (high $\eta$). Therefore, we reject the objects with $\eta < -0.15$ and $\eta > 0.2$  from the SFI++ dataset. We then iteratively reject the points which are not within $3.5\sigma$ of the inferred TF relation in the magnitude. Finally, we compare the peculiar velocity predicted using bulk flow parameters inferred using the $\chi^2$ minimization method (described in section \ref{ssec:chisquared}) to the reported peculiar velocities in the SFI++ dataset. We reject the $3.5 \sigma$ outliers (17 objects) from this comparison. 

For fitting the bulk flow parameters, we use both the galaxy and the group catalogues from the SFI++ dataset. Therefore, we remove the duplicates from the galaxy catalogue in the group catalogue. We also reject the groups for which all the corresponding galaxies in the dataset are rejected during one of the cuts described in the earlier paragraph. After these cuts, we are left with a total of $1996$ field galaxies and $599$ groups (containing $1167$ galaxies). The characteristic depth of the field galaxy sample was found to be $40~h^{-1}$ Mpc and that of the group sample is $22~h^{-1}$ Mpc.

\subsubsection{2MTF}

The 2MTF survey  \citep{2MTF, 2mtf_data} contains TF data for $2062$ galaxies in the nearby Universe. It is restricted to distances $< 100~h^{-1}$ Mpc. For objects which are in both SFI++ and 2MTF catalogues, we use the objects in the SFI++ catalogue. This is because the TF relation in the I-band (used by SFI++) is found to have a smaller scatter compared to the TF relation in the infrared (employed in the 2MTF) frequency. It is reflected in our results of scatter as found for the SFI++ catalogue and the 2MTF (See Table \ref{tbl:tf_parameters}). To remove duplicates, we cross-match the galaxies by considering all 2MTF galaxies with an angular separation of $\le 20$ arcseconds and a redshift difference $| \Delta cz| < 150$ km s$^{-1}$ of SFI++ galaxies. We find $384$ galaxies that are in both catalogues, and we remove these from the 2MTF sample. We also include only galaxies that are in the 2M++ region, which removes another $22$ galaxies. The 2MTF survey provides galaxy magnitudes in the $H$, $J$ and $K$ bands. For the purposes of this paper, we only use the $K$ band magnitudes. As with the SFI++ data, we observe a deviation from the inferred linear relationship at the faint and bright ends of the sample. We therefore keep only galaxies with $-0.1 < \eta < 0.2$. We then fit the Tully-Fisher relationship by using the redshift-space distance and iteratively exclude the $3.5\sigma$ outliers. The final sample has a total of $1247$ galaxies. The characteristic depth of the 2MTF sample is $21~h^{-1}$ Mpc.
\section{Density and velocity field reconstruction}
\label{sec:reconstruction}

In this section, we present details on the density and velocity reconstruction that we use for predicting the peculiar velocities. In section \ref{ssec:2mpp_data}, we describe the 2M++ redshift compilation, which has been used in our reconstruction. In section \ref{ssec:reconstruction_scheme}, we present details of the reconstruction scheme used.

\subsection{2M++ galaxy redshift compilation}\label{ssec:2mpp_data}

Peculiar velocities are sourced by the density fields on large scales. Therefore to study peculiar velocities, we require our galaxy catalogue to have a large sky coverage and be as deep as possible. With this as a goal, the 2M++ compilation of galaxy redshifts was constructed in \citet{2Mpp_paper}. The 2M++ redshifts are derived from the 2MASS redshift survey (2MRS) \citep{2mass1}, 6dF galaxy redshift survey-DR3 \citep{6df} and the Sloan Digital Sky Survey (SDSS) Data Release 7 \citep{sdss_dr7}. The apparent K-band magnitude was corrected by taking into account Galactic extinction, $k$-corrections, evolution and surface brightness dimming. The Zone of Avoidance  (ZoA) due to the Galactic Plane is masked in the process. The resulting catalogue consists of a total of $69160$ galaxies. The catalogue was found to be highly complete up to a distance of $200\ h^{-1}$ Mpc (or $K < 12.5$) for the region covered by the 6dF and SDSS and up to $125\ h^{-1}$ Mpc (or $K < 11.5$) for the region that is not covered by these surveys. Hereafter, `2M++ volume/region' is restricted to less than $200\ h^{-1}$ Mpc for the region in the 2M++ catalogue which is covered by SDSS and 6dF survey and to less than $125\ h^{-1}$ Mpc for the region covered only by 2MRS.

In \citet{Carrick_et_al}, the ZoA was filled by ``cloning'' galaxies above and below the plane. We elaborate on the reconstruction process in section \ref{ssec:reconstruction_scheme}. For further details on the 2M++ catalogue, see \citet{2Mpp_paper} and the references therein. 

\subsection{Reconstruction scheme}
\label{ssec:reconstruction_scheme}

In \citet{Carrick_et_al}, the density field was reconstructed with an iterative scheme modelled on \citet{yahil_et_al}. We use the luminosity-weighted density field from \citet{Carrick_et_al} in this work. 
A luminosity weight was assigned to each galaxy in the 2M++ catalogue after fitting the luminosity function with a Schechter function. Galaxy bias depends on luminosity: \citet{westover} found that
\begin{equation}
    \frac{b}{b_{*}} = (0.73 \pm 0.07) + (0.24 \pm 0.04)\frac{L}{L_{*}}\,,
\end{equation}
where $b_{*}$ is the bias of an $L_{*}$ galaxy. This luminosity-dependent bias function was used to normalize the density contrast to a uniform  $b_{*}$ at all radii.

\begin{figure*}
    \centering
    \includegraphics[width=\linewidth]{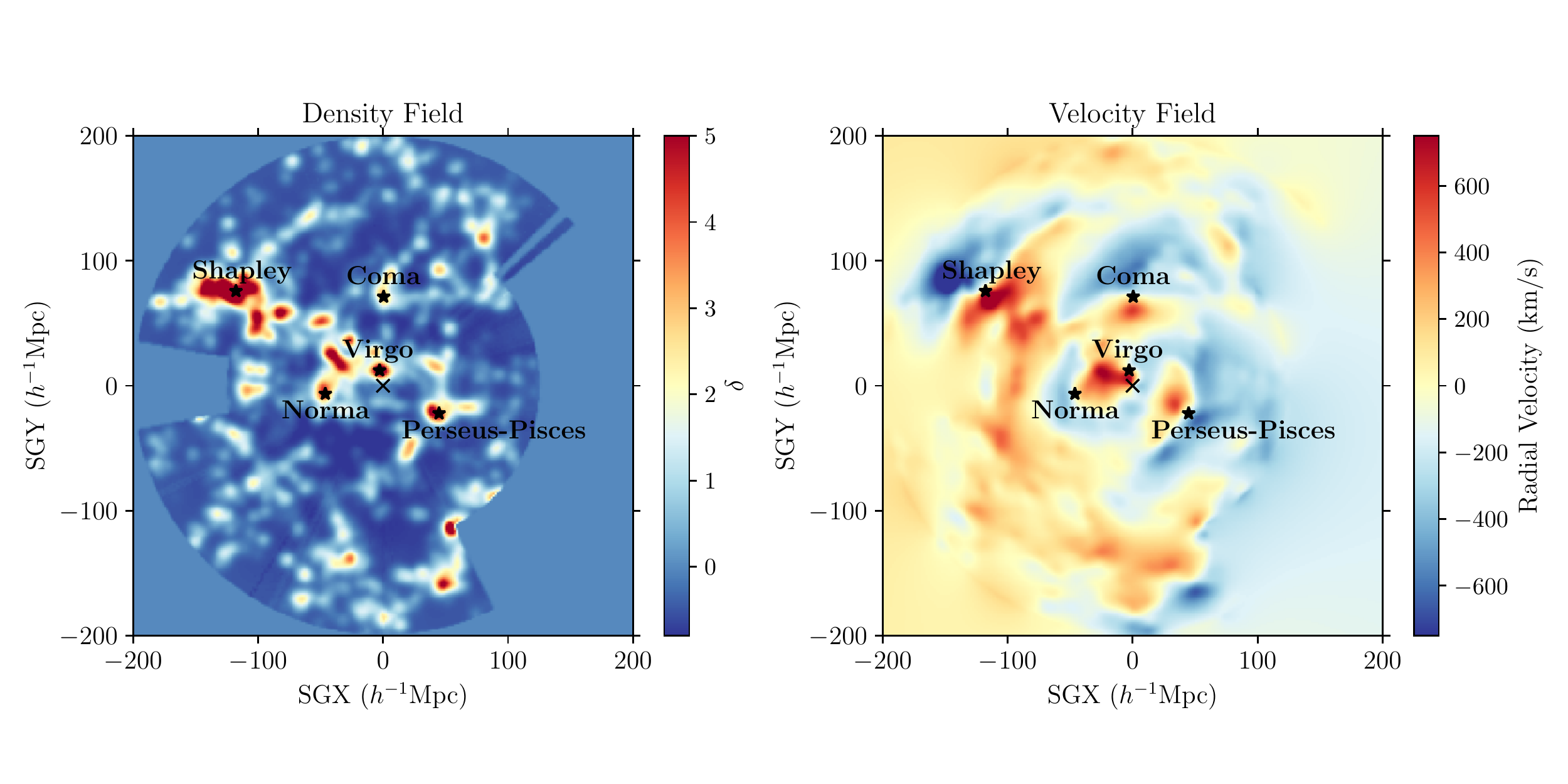}
    \caption{The reconstructed luminosity-weighted density ($\delta_g$) and the radial velocity field in the Supergalactic Plane smoothed with a Gaussian filter of size $4~h^{-1}$ Mpc. The location of the prominent superclusters, namely, Shapley, Coma, Virgo, Norma and Perseus-Pisces are shown with a black star. The Local Group is at the origin and is denoted with a black cross.}
    \label{fig:supergalactic}
\end{figure*}

Finally, the mapping from the redshift data of 2M++ to comoving coordinates is done using an iterative scheme. First, the galaxies are grouped using the `Friends of friends' algorithm \citep{fof}. Then, galaxies are initially placed at the comoving distance corresponding to its redshift. Then, the luminosity-weighted density field is calculated and  smoothed using a Gaussian filter at $4\ h^{-1}$ Mpc. From this density field, the peculiar velocity is calculated using linear theory for each object. The comoving coordinates in the next iteration are then corrected for using this peculiar velocity prediction. This process is repeated, slowly increasing $\beta$ from $\beta=0$ to $\beta=1$. For more details on this reconstruction procedure, refer to \cite{Carrick_et_al}. The reconstructed density and the radial velocity in the supergalactic plane is shown in Figure \ref{fig:supergalactic}.

\section{Comparing Predicted and Observed Peculiar Velocities}\label{sec:compare}

We want to compare the reconstructed velocity field to the observations of the peculiar velocity catalogues. To do this, we fit for $\beta = f/b$ and a coherent residual bulk flow velocity $\mvec{V}_{\text{ext}}$, which may arise from the large-scale structures outside of the 2M++ survey area. Note that the published velocity field\footnote{Available at \href{https://cosmicflows.iap.fr}{https://cosmicflows.iap.fr}} of \citet{Carrick_et_al} uses $\beta = 0.43$ with the $\mvec{V}_{\text{ext}}=(89,-131,17)$ km s$^{-1}$ as given in that paper. Here we refit both $\beta$ and $\mvec{V}_{\text{ext}}$.

The approaches to measuring peculiar velocities via distance indicators often come in two variants, the so-called forward and the inverse approaches. In the \textit{forward approach}, one predicts a distance-dependent quantity (e.g.\ magnitude) as a function of a distance-independent quantity (e.g.\ velocity width). In the \textit{inverse approach} \citep{aaronson82}, one predicts the distance-independent quantity as a function of a distance-dependent quantity. Note, however, that some distance indicators, such as Type Ia supernovae, only have a ``forward'' method. In addition to the above distinction between the forward and inverse methods, there is another distinction that is generally made between the different approaches. Predicting the peculiar velocity requires an \textit{a priori} `best estimate' for the position of the observed galaxies. One can use, for example, the Tully-Fisher relations to assign an \textit{a priori} best estimate of the distance to a galaxy. Alternately, one can use the redshift as the \textit{a priori} best estimate of the distance. The former has been called the \textit{Method I} and the latter, \textit{Method II} in the literature \citep{pec_vel_review}. 

Each combination of distance-indicator method and Method I/II are subject to different biases which arise due to selection effects and density inhomogeneities. However, biases are lower for some combinations: in particular, an inverse distance indicator combined with Method II is insensitive to Malmquist biases arising from the scatter in the distance indicator\footnote{There remains a weak Malmquist-like bias due to the scatter in the flow model used to assign a distance given a redshift \citep{KaiHud15b} but this is much smaller than the one due to the scatter in the distance indicator.}, whereas a forward distance indicator combined with Method I is less sensitive to many selection effects.

We use two different methods for our peculiar velocity analysis: a simple $\chi^2$ minimization technique and a forward likelihood method.

\subsection{$\chi^2$ minimization}\label{ssec:chisquared}

In the first approach, we compare the observed redshift to the predicted redshift of a galaxy by assuming it is at the distance reported in the peculiar velocity survey. This is therefore a \textit{Forward-Method I} approach. This approach suffers from Malmquist bias \citep{pec_vel_review}. In Section~\ref{ssec:forward}, we correct for the Malmquist bias by integrating the measured inhomogeneities along the line-of-sight. It is difficult to correct for it in a simple $\chi^2$ fitting method used in this subsection. Because of this bias, the inferred value of $\beta$ is biased high in this approach. Nevertheless, we use this method because of its interpretability and to check consistency. 

The predicted redshift for a galaxy is dependent on the flow model and the reconstructed velocity. That is, $z_{\text{pred}} \equiv z_{ \text{pred}}(\mvec{r},\mvec{v},\mvec{V}_{\text{ext}}, \beta)$. The dependence of $z_{\text{pred}}$ on these quantities is given as 
\begin{equation}\label{eqn:z_pred}
    1 + z_{\text{pred}} = \bigg(1 + z_{\text{cos}}(r)\bigg)\bigg(1 + \frac{1}{c}(\beta \mvec{v} + \mvec{V}_{\text{ext}})\cdot\mvec{\hat{r}}\bigg)\,,
\end{equation}
where $r$ is obtained by taking the distance as being equal to the reported distance in the peculiar velocity catalog and $\mvec{v}$ is the velocity predicted from our reconstruction. In what follows, we do not explicitly show the dependence of $z_{\text{pred}}$ on the reconstructed velocity and the flow model. For the cosmological redshift, $z_{\text{cos}}$, we use a second order approximation \citep{Peebles_book}, 
\begin{equation}\label{eqn:z_cos_relation}
    z_{\text{cos}}(r) = \frac{1}{1 + q_0} \bigg[1 - \sqrt{1 - \bigg(\frac{2 H_0 r}{c}\bigg)(1 + q_0))} \bigg]\,,
\end{equation}
where, $q_0$ is the deceleration parameter, which can be related to the cosmological parameters, $\Omega_m$ and $\Omega_{\Lambda}$ as, $q_0 = \frac{\Omega_m}{2} - \Omega_{\Lambda}$. This approximation is accurate to better than 2 km s$^{-1}$ in $cz$ for $z < 0.05$.  The relation between $z_{\text{cos}}$ and the comoving distance $r$ is also robust to the adopted value of $\Omega_m$: a $5\%$ difference in $\Omega_m$ results in $< 0.5\%$ difference in the $z_{\text{cos}}$.

Given the parameters $(\beta, \mvec{V}_{\text{ext}})$ and a reconstructed velocity field $\mvec{v}$, the discrepancy between the observations and the predictions is given by,
\begin{equation}\label{eqn:chi_sq}
	\chi^2(\beta, \mvec{V}_{\text{ext}}) = \sum_{i=1}^{N_{\text{gal}}} \frac{(cz_{\text{obs}} - cz_{\text{pred}})^2}{\sigma^2_d + \sigma^2_v}\,,
\end{equation}
where $\sigma_v$ is the additional uncertainty in modelling the velocity field and $\sigma_d$ is the error on the distance estimate converted to the units of km s$^{-1}$. The predicted redshift, $z_{\text{pred}}$ is obtained using Equation \eqref{eqn:z_pred} by assuming that the tracer is at the radial position reported in the peculiar velocity catalogue. Unless mentioned otherwise, throughout this work, we fix $\sigma_v = 150\; \textrm{km s}^{-1}$. This value was obtained in \cite{Carrick_et_al} by comparing the linear theory predictions with the observed velocities of halos in N-body simulations. However, changing this value (or fitting it as an additional parameter) does not change the results.

We minimize the $\chi^2$ given in Equation \eqref{eqn:chi_sq} with respect to $\beta$ and $\mvec{V}_{\text{ext}}$ to infer the best-fit flow model. 

\begin{figure*}
  \centering
  \includegraphics[width=0.95\linewidth]{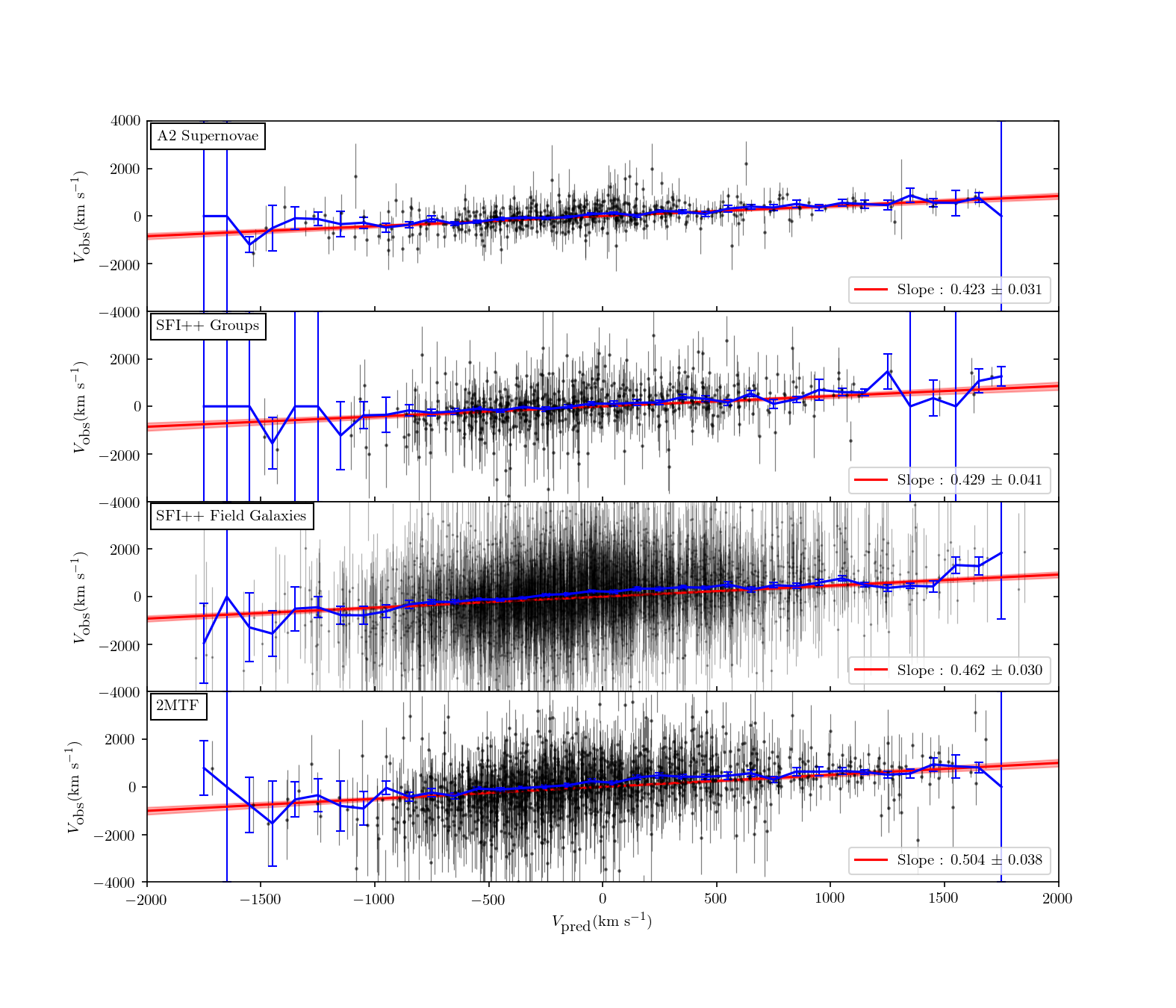}
\caption{The predicted velocity ($V_{\textrm{pred}}$) vs the observed velocity ($V_{\textrm{obs}}$) for objects in the peculiar velocity catalogues: the A2 supernovae, SFI++ groups, SFI++ field galaxies and 2MTF. The predicted velocity is scaled to $\beta = 1$. The fitted slope therefore gives an estimate for $\beta$, although this will be biased somewhat high due to inhomogeneous Malmquist bias (see text for details). The red solid line is the best fitted line and the shaded area is the corresponding $1\sigma$ error. The blue lines are the weighted average of $V_{\text{obs}}$ in bins of width $100$ km/s in $V_{\text{pred}}$.}
\label{fig:V_obs_V_pred}
\end{figure*}

\subsection{Forward Likelihood}\label{ssec:forward}

As mentioned in the previous section, the \textit{Forward-Method I} approach is affected by inhomogeneous Malmquist bias. \citet{pike_hudson} introduced an approach to take care of these difficulties. We call this approach  \textit{Forward likelihood}. A virtue of this method is that we can include any distance indicator data in this method.

Like the $\chi^2$ minimization method introduced in section \ref{ssec:chisquared}, the difference in the observed and predicted redshifts are minimized in this approach. To correct the inhomogeneous Malmquist bias, we need to take the inhomogeneities along the line of sight into account. Also, the measured distances have a lognormal uncertainty\footnote{Although using a Gaussian distribution changes the inferred value of $\beta$ by $< 0.35\sigma$}. To correct for these deficiencies, we assume that the radial distribution is given as,
\begin{equation}\label{eqn:r_prior}
	\mathcal{P}(r) = \frac{r^2 \exp\bigg(-\frac{[\mu(r) - \mu(d)]^2}{2\sigma_{\mu}^2}\bigg) [1 + \delta_g(\mvec{r})]}{\int_0^{\infty} \diffop r^{\prime} r^{\prime 2} \exp\bigg(-\frac{[\mu(r^{\prime}) - \mu(d)]^2}{2\sigma_{\mu}^2}\bigg) [1 + \delta_g(\mvec{r^{\prime}})]}\;,
\end{equation}
where $d$ is the distance reported in the peculiar velocity survey and $\delta_g$ is the overdensity in the galaxy field. The distances are converted to distance modulus using the formula $\mu(r) = 5\log_{10}(r/10$ pc). $\sigma_{\mu}$ is the error in the distance modulus of the tracer. As a proxy for the galaxy field, the luminosity weighted density was used. For the distance estimates which have already been corrected for homogeneous Malmquist bias, we drop the $r^2$ term from the prior. Instead, to correct for possible scale errors in the reported distance, we marginalize over a nuisance parameter, $\tilde{h}$, which rescales the reported distance
\begin{equation}\label{eqn:r_prior_htilde}
	\mathcal{P}(r|\tilde{h}) = \frac{1}{\mathcal{N}(\tilde{h})}\exp\bigg(-\frac{[\mu(r) - \mu(\tilde{h}d)]^2}{2\sigma_{\mu}^2}\bigg) [1 + \delta_g(\mvec{r})]\;,
\end{equation}
where $\mathcal{N}(\tilde{h})$ is the normalization term that depends on $\tilde{h}$. To account for the errors that arise because of the triple-valued regions and inhomogeneities along the line of sight, the likelihood is marginalized over the above radial distribution. The likelihood, $P(z_{\text{obs}}|\mvec{v}, \mvec{V}_{\text{ext}}, \beta)$, can therefore be written as
\begin{equation}\label{eqn:fwd_lkl}
	\mathcal{P}(z_{\text{obs}}|\mvec{v}, \mvec{V}_{\text{ext}}, \beta) = \int_0^{\infty} dr \mathcal{P}(z_{\text{obs}}| r, \mvec{v}, \mvec{V}_{\text{ext}}, \beta) \mathcal{P}(r),
\end{equation}
where
\begin{equation}\label{eqn:conditional_prob_z}
	\mathcal{P}(z_{\text{obs}}|r, \mvec{v}, \mvec{V}_{\text{ext}}, \beta)  = \frac{1}{\sqrt{2\pi \sigma^2_v}}\exp\bigg(-\frac{(cz_{\text{obs}} - c z_{\text{pred}})^2}{2\sigma^2_v}\bigg)\,,
\end{equation}
and $\mathcal{P}(r)$ is given by Equation~\eqref{eqn:r_prior} and $z_{\text{pred}} \equiv z_{\text{pred}}(r, \mvec{v}, \mvec{V}_{\text{ext}}, \beta)$ as given in Equation~\eqref{eqn:z_pred}.

We infer the flow model, $\{\beta$, $\mvec{V}_{\text{ext}}\}$ by sampling from the following posterior distribution,  
\begin{equation}\label{eqn:fwd_posterior}
	\mathcal{P}(\mvec{V}_{\text{ext}}, \beta| \mvec{v}, z_{\text{obs}}) = \frac{\mathcal{P}(z_{\text{obs}}|\mvec{v}, \mvec{V}_{\text{ext}}, \beta) \mathcal{P}(\mvec{V}_{\text{ext}}, \beta)}{\mathcal{P}(z_{\text{obs}})}\,.
\end{equation}
Assuming a uniform prior on $V_{\text{ext}}$, $\beta$ and ignoring the denominator in Equation~\eqref{eqn:fwd_posterior} as it does not depend on the parameters of interest, the posterior turns out to have the same functional form as the likelihood. 

For the dataset of all galaxies, $\{z_i\}$, assuming independent probabilities, we maximize the joint posterior, which is given by 
\begin{equation}
	\mathcal{P}(\mvec{V}_{\text{ext}}, \beta,\tilde{h}| \{ z_i\}) \propto \prod_i \mathcal{P}(z_i | \mvec{V}_{\text{ext}}, \beta)\,.
\end{equation}

The results from the forward likelihood fit are presented in Section \ref{sssec:fwd_lkl_results}.

\subsubsection{Jointly inferring distances and flow model with a modified Forward likelihood method}
\label{sssec:fwd_lkl_joint}

Measuring distances usually requires a calibration step for the distance indicator relationship. In this section, we introduce a method to jointly calibrate the distance indicator relationship while fitting for the flow model. To fit the LOSS and Foundation supernovae data and the field galaxies sample of the SFI++ catalogue, we modify the forward likelihood method to jointly fit for both the flow model and the parameters of the distance indicator. For the SALT2 model, the distance is a function of the global parameters, $\alpha$, $\mathcal{B}$, $M$ and $\sigma_{\textrm{int}}$. We jointly denote these parameters with $\Theta_{\textrm{SN}}$. Similarly, for the Tully-Fisher relationship, the distances depend on the TF parameters, $\Theta_{\textrm{TF}} = \{ a_{\text{TF}}, b_{\text{TF}}, \sigma_{\textrm{int}}\}$. In order to jointly fit the parameters of distance indicator and the flow model, we therefore fit for these global parameters in addition to the flow model. In this approach, the Equation~\eqref{eqn:r_prior} is modified to 
\begin{equation}\label{eqn:r_prior_sn}
    \mathcal{P}(r|\Theta) = \frac{1}{\mathcal{N}(\Theta)}r^2 \exp\bigg(-\frac{[\mu(r)-\mu(\Theta)]^2}{2\sigma_{\mu}^2(\Theta)}\bigg) [1 + \delta_g(\mvec{r})]\;,
\end{equation}
where $\mu(\Theta)$ is obtained from equation \eqref{eqn:tripp} or \eqref{eqn:tf_distance} and $\sigma_{\mu}$ is obtained by adding in quadrature the intrinsic scatter and the measurement uncertainty. Here, $\Theta$ stands for either $\Theta_{\text{SN}}$ or $\Theta_{\text{TF}}$. Note that we have added back the volume term which corrects fro the homogeneous Malmquist bias. $\mathcal{N}(\Theta)$ is the normalization term that depends on $\Theta$. Using Bayes' Theorem as in the usual approach, we can then write the joint posterior for $\Theta, \mvec{V}_{\textrm{ext}}, \beta$. We sample from this posterior to infer the parameters. The results for fitting the supernovae data in using this method are presented in section \ref{sec:results}. 

To sample from the posterior distribution of this section, we used the MCMC package \texttt{emcee} \citep{emcee_1, emcee_2}. The autocorrelation time for the MCMC chains is $\mathcal{O}(10$-$20)$\footnote{We note that finding the autocorrelation of an ensemble sampler is not trivial as the walkers are not independent. To get our estimate, we calculated the autocorrelation for each walker and then average over them. This has been suggested in \url{https://emcee.readthedocs.io/en/stable/tutorials/autocorr/}}. This gives an effective sample size of $\sim 1000$.

\section{Results}\label{sec:results}

In this section, we will present the results of the comparison between the predicted and the measured peculiar velocities. In Section~\ref{ssec:velocity_analysis_results}, we present our peculiar velocity analysis of the different catalogues. In Section \ref{ssec:constraints}, we present the constraints on the cosmological parameters and the external bulk flow.

\subsection{Peculiar velocity analysis with different catalogues}\label{ssec:velocity_analysis_results}

In this section, we present the results of analysis of the different catalogues we use in our peculiar velocity analysis. First, we analyse these catalogues using the $\chi^2$-minimization method presented in section \ref{sssec:chi_squared_results}. In section \ref{sssec:fwd_lkl_results}, we present the analysis of the same catalogues using the forward likelihood method.

\subsubsection{$\chi^2$ minimization}\label{sssec:chi_squared_results}
\begin{table*}
  \centering
  \caption{Results of the $\chi^2$ minimization with the different catalogues. Note that the $\chi^2$ method is affected by inhomogeneous Malmquist bias. We correct for the IHM using the forward likelihood method. Forward likelihood result is presented in Table \ref{tbl:fwd_lkl}.}
  \begin{tabular}{l|c|c|c|c|c}
  \hline
  \hline
     Sample & $\beta$ & $|\mvec{V}_{\textrm{ext}}|$(km/s) & $l$(deg) & $b$(deg) & $\chi^2$/d.o.f\\
    \hline
    \hline
   A1 & $0.445 \pm 0.042$ & $130 \pm 37$ & $314 \pm 29$ & $26\pm 17$  & $0.901$\\   CSP-DR3 & $0.588 \pm 0.092$ & $231 \pm 63$ & $14 \pm 41$ & $-50 \pm 18$ & $0.950$\\
    LOSS & $0.483 \pm 0.077$ & $264 \pm 100$ & $282 \pm 42$ & $-24 \pm 16$ & $1.005$\\
    Foundation & $0.389 \pm 0.060$ & $375 \pm 64$ & $251 \pm 10$ & $18 \pm 7$  & $1.032$\\
    \textbf{A2} & $\boldsymbol{0.439 \pm 0.033}$ & $\boldsymbol{132 \pm 30}$ & $\boldsymbol{285 \pm 17}$ & $\boldsymbol{16 \pm 13}$  & $\boldsymbol{0.768}$\\
    SFI++ Groups & $0.431 \pm 0.040$ & $184 \pm 41$ & $282 \pm 22$ & $23 \pm 14$ & $0.839$\\
    SFI++ Field Galaxies & $0.458 \pm 0.031$ & $192 \pm 30$ & $283 \pm 11$ & $4 \pm 8$  & $0.734$\\
    2MTF & $0.504 \pm 0.041$ & $190 \pm 36$ & $285 \pm 16$ & $17 \pm 11$  & $0.938$\\
    \textbf{Combined} & $\boldsymbol{0.457 \pm 0.016}$ & $\boldsymbol{163 \pm 17}$ & $\boldsymbol{283 \pm 8}$ & $\boldsymbol{15 \pm 6}$  & $\boldsymbol{0.803}$\\
    \hline
  \end{tabular}
  \label{tbl:chi_squared}
\end{table*}
While the $\chi^2$ minimization method is affected by the inhomogeneous Malmquist bias, it is advantageous to get interpretable results. We present the results of the $\chi^2$-minimization method in Table \ref{tbl:chi_squared}.  For each sub-sample, we infer the external bulk flow velocity, $\mvec{V}_{\textrm{ext}}$, its direction in the galactic coordinates, $l$ and $b$. 
We also infer the velocity rescaling factor for the predicted velocity from reconstruction. Note that this rescaling factor is equal to $\beta = f/b$. In Table \ref{tbl:chi_squared}, we also report the value of the $\chi^2$ over the number of degrees of freedom. In this section, for the SFI++ catalogue, we use the distance as reported in the catalogue. For the supernovae samples, we use a variant of the $\chi^2$ minimization method where we also fit for the intrinsic scatter. For the LOSS and the Foundation sample, we fit for the light curve parameters in addition to the flow model parameters. 

In Figure \ref{fig:V_obs_V_pred}, we compare the predicted peculiar velocities to the observations from the peculiar velocity surveys. In the $\chi^2$ minimization method, the difference between the two is minimized by fitting for the flow model. The observed peculiar velocities usually have a large uncertainty. Nonetheless, when taken together, the trend is clearly visible. We also show the results of the $\chi^2$ fitting method in the plot. We plot the predicted velocities from the reconstruction against the observed velocities and fit for the slope. This slope roughly corresponds to the value of $\beta$. However, the obtained value is biased high due to inhomogeneous Malmquist bias. One can also observe this by comparing the value of $\beta$ as found in Table \ref{tbl:chi_squared} and Table \ref{tbl:fwd_lkl}.

\subsubsection{Forward likelihood}\label{sssec:fwd_lkl_results}

\begin{table*}
  \centering
  \caption{Results of forward likelihood analysis for different peculiar velocity datasets. For the A2 and the combined results, we jointly fit the flow model parameters and the global parameters of the each sample.}
  \begin{tabular}{l|c|c|c|c|c}
  \hline
  \hline
     Sample & $\beta$ & $f\sigma_{8,\textrm{lin}}$ & $|\mvec{V}_{\textrm{ext}}|$(km/s) & $l$(deg) & $b$(deg)\\
    \hline
    \hline 
   A1 & $0.421 \pm 0.030$ & $0.396\pm0.030$ & $156^{+32}_{-31}$ & $310^{+13}_{-13}$ & $7^{+9}_{-9}$\\[0.1cm]
    CSP-DR3 & $0.483 \pm 0.103$ & $0.469 \pm 0.097$ & $217^{+55}_{-55}$ & $19^{+32}_{-31}$ & $-43^{+16}_{-17}$\\[0.1cm]
    LOSS & $0.490 \pm 0.079$ & $0.456 \pm 0.073$ & $150^{+65}_{-65}$ & $282^{+105}_{-99}$ & $-26^{+27}_{-27}$\\[0.1cm]
    Foundation & $0.345 \pm 0.064$ & $0.331 \pm 0.062$ & $314^{+54}_{-55}$ & $249^{+12}_{-12}$ & $17^{+8}_{-8}$\\[0.1cm]
    \textbf{A2} & $\boldsymbol{0.408 \pm 0.025}$ & $\boldsymbol{0.385 \pm 0.027}$ & $\boldsymbol{135^{+24}_{-24}}$ & $\boldsymbol{300^{+11}_{-11}}$ & $\boldsymbol{-3^{+8}_{-8}}$\\[0.1cm]
    SFI++ groups & $0.411\pm0.027$& $0.385 \pm 0.030$ & $174^{+31}_{-30}$ & $292^{+10}_{-10}$ & $2^{+7}_{-7}$ \\[0.1cm]
    SFI++ field galaxies & $0.411 \pm 0.020$ & $0.385 \pm 0.022$ & $181^{+20}_{-21}$ & $291^{+10}_{-10}$ & $14^{+6}_{-6}$ \\[0.1cm]
    2MTF & $0.483 \pm 0.022$ & $0.444 \pm 0.025$ & $177^{+19}_{-19}$ & $304^{+7}_{-7}$ & $-2^{+5}_{-5}$ \\[0.1cm]
    \textbf{All combined} & $\boldsymbol{0.428 \pm 0.012}$ & $\boldsymbol{\fsigmainv}$ & $\boldsymbol{\vextinv}$ &  $\boldsymbol{301^{+4}_{-4}}$ & $\boldsymbol{0^{+3}_{-3}}$ \\[0.1cm]
    \hline
  \end{tabular}
  \label{tbl:fwd_lkl}
\end{table*}

We also analysed the peculiar velocity samples using the forward likelihood method of Section~\ref{ssec:forward}. The result of this analysis is presented in Table~\ref{tbl:fwd_lkl}. For the analysis in this section, wherever possible, we use the modified forward likelihood method, presented in Section~\ref{sssec:fwd_lkl_joint} to jointly fit for the distance indicator parameters and the flow model. For the Foundation and the LOSS SNe samples and the Tully-Fisher galaxy samples, we fit the parameters of the distance indicator relation and the flow model for the sample. In this method, the likelihood is still given by Equation~\eqref{eqn:fwd_lkl} but Equation~\eqref{eqn:r_prior} is modified to Equation~\eqref{eqn:r_prior_sn}. We jointly fit $\beta, \mvec{V}_{\text{ext}}, M, \alpha, \mathcal{B}$ and $\sigma_{\textrm{int}}$ for the supernovae samples and $\beta, \mvec{V}_{\text{ext}}, a_{\text{TF}}, b_{\text{TF}}, \sigma_{\text{int}}$ for the Tully-Fisher samples using this modified version of forward likelihood. Similarly, for the CSP-DR3 sample, we also fit the intrinsic scatter. The results of these fits and a comparison of the intrinsic scatter for the LOSS, CSP and Foundation sample is presented in Table \ref{tbl:light_curve}.

\begin{table*}
  \centering
  \caption{Light curve parameters and intrinsic scatter inferred using the modified forward likelihood analysis for the LOSS, Foundation and the CSP samples}
  \begin{tabular}{l|c|c|c|c}
  \hline
  \hline
     Sample & $M + 5\log_{10}(h)$ & $\alpha$ & $\mathcal{B}$ & $\sigma_{\textrm{int}}$ (mag) \\
    \hline
    \hline
   LOSS & $-18.191 \pm 0.022$ & $0.121 \pm 0.019$ & $3.53 \pm 0.16$ & $0.125 \pm 0.017$\\
    Foundation & $-18.558 \pm 0.011$ & $0.140 \pm 0.011$ & $2.78 \pm 0.12$ & $0.082 \pm 0.010$\\
    CSP-DR3 & --- & --- & --- & $0.062 \pm 0.018$ \\
    \hline
  \end{tabular}
  \label{tbl:light_curve}
\end{table*}

\subsection{Constraining $f\sigma_8$ and the bulk flow}\label{ssec:constraints}

In this section we present the results of inferring the cosmological parameter, $f\sigma_8$, and the bulk flow using the forward likelihood method. Note that while inferring the flow model with multiple catalogues, we jointly fit the distance indicator parameters of each peculiar velocity catalogue and the flow model parameters.

\subsubsection{Constraint on $f\sigma_8$}\label{sssec:fsigma8_results}

Using the forward likelihood method of Section~\ref{ssec:forward}, we inferred the parameter $\beta = f/b$ for the reconstructed velocity field. The relation between $\beta$ and the factor $f\sigma_8$ is given as, $f\sigma_8 = \beta \sigma^g_8$, where $\sigma^g_8$ is the root mean squared fluctuation in the galaxy field. \cite{Carrick_et_al} found the value of $\sigma^g_8$ to be $0.99 \pm 0.04$. To convert our constraints on $\beta$ to the constraints on $f \sigma_8$, we use this value of $\sigma^g_8$. 

\begin{figure}
\centering
  \includegraphics[width=\linewidth]{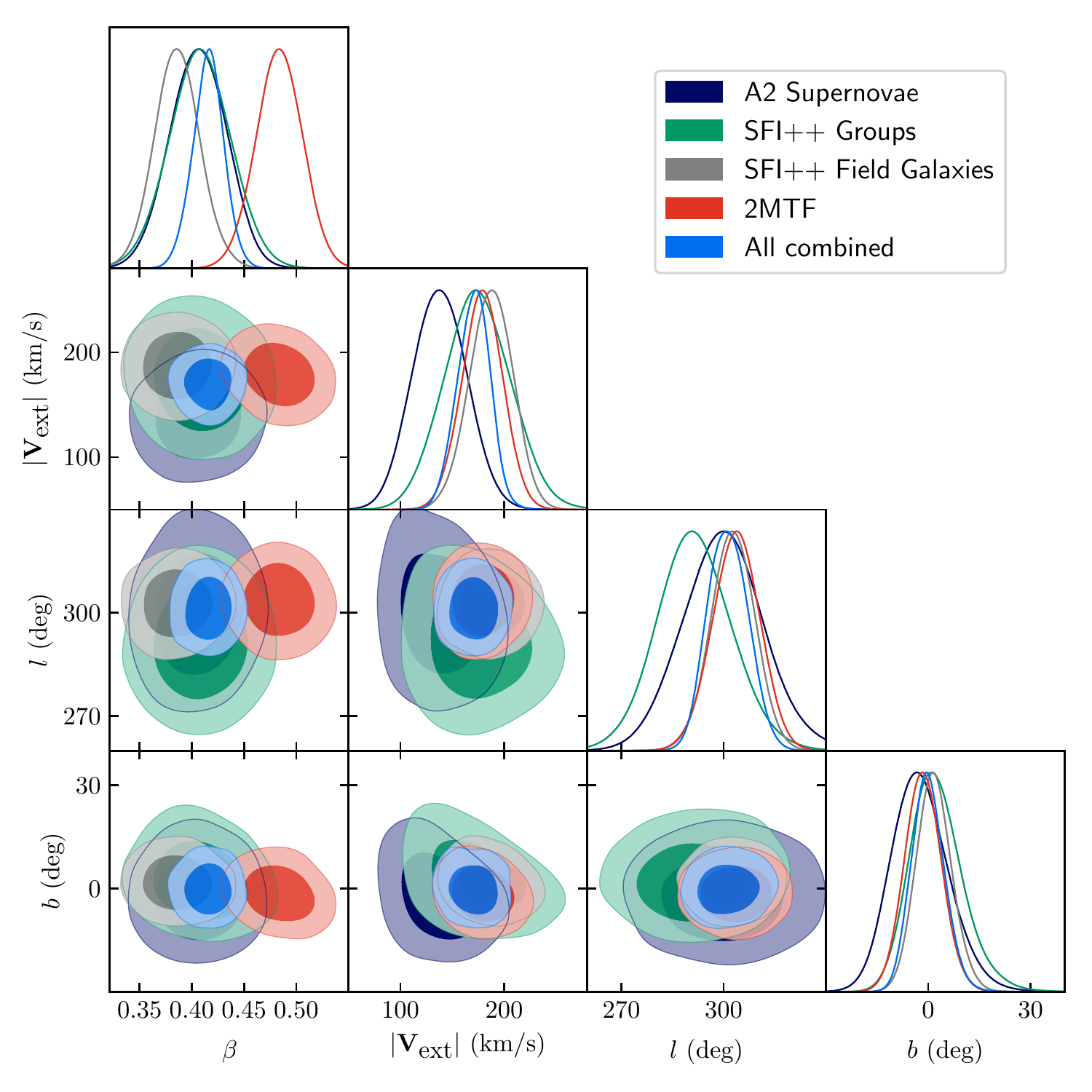}
\caption{The results of forward likelihood inference with our reconstruction scheme. The numerical values are presented in Table \ref{tbl:fwd_lkl}. The panels show the two dimensional marginal posteriors for $\beta, |\mvec{V}_{\textrm{ext}}|, l, b$. The different samples corresponds to the results obtained from the taking the different datasets. The `combined' dataset is obtained by combining the A2 supernovae, SFI++ field galaxies, SFI++ groups and 2MTF samples. The dark and light shaded regions correspond to the $68\%$ and $95\%$ confidence intervals respectively}
\label{fig:corner}
\end{figure}

It should be noted however, the value of $\sigma_8$ inferred from peculiar velocities is sensitive to the non-linear evolution of structures. To compare with values of $\sigma_8$ inferred at high redshifts, we need to correct for the non-linear evolution. This is done using the recipe of \citet{sigma8_correction}. This linearized value is denoted by $f \sigma_{8, \textrm{lin}}$. We assume $\Omega_m = 0.3$ to convert the constraint on $f\sigma_8$ into the constraint on $\sigma_8$. This value is then converted into the linearized value. The result for $f \sigma_{8,\textrm{lin}}$ as inferred from the two reconstruction schemes and the different datasets is presented in Table \ref{tbl:fwd_lkl} and in Figure \ref{fig:corner}. We find consistent results from the different datasets. The value of $f\sigma_{8,\textrm{lin}}$ inferred by comparing the combined dataset of the A2 supernovae, 2MTF and the SFI++ to the predictions of our reconstruction is $\fsigmainv$.

\subsubsection{Bulk Flow}

We also infer the external bulk flow in the forward likelihood method. The bulk flow may be thought of as the coherent flow in the reconstructed volume. Comparison with our reconstruction yields an external bulk flow of magnitude $\vextinv$ km s$^{-1}$ in the direction $l = \lextinv, b = \bextinv$. We also compare the reconstructed bulk flow centred on the Local Group at an effective radius of $40~h^{-1}$ Mpc. At this scale, we find a bulk flow of $\vbulkinv$ km s$^{-1}$ in the direction $l=\lbulkinv$, $b = \bbulkinv$. To obtain this flow, we added the external flow to the velocity obtained by smoothing the reconstructed velocity flow at $R=40~h^{-1}$ Mpc with a Gaussian filter. We compare our results for the bulk flow at $40~h^{-1}$ Mpc with other results from the literature in section \ref{ssec:compare_bulk_flow}.

\section{Discussion}\label{sec:discussion}

In this section, we compare our inferred value of $f\sigma_8$ and the bulk flow to that of other results in the literature and also discuss the prospects for the future. 

\subsection{Comparison with the literature}

In Section~\ref{ssec:compare_fsigma8}, we compare our results to other results of $f\sigma_8$ based on a variety of cosmological probes. In section \ref{ssec:compare_bulk_flow}, we compare our results for the bulk flow to $\Lambda$CDM prediction and to other results in the literature. 

\subsubsection{Comparison of the value of $f \sigma_8$}\label{ssec:compare_fsigma8}

\begin{figure*}
    \centering
    \includegraphics[width=\linewidth]{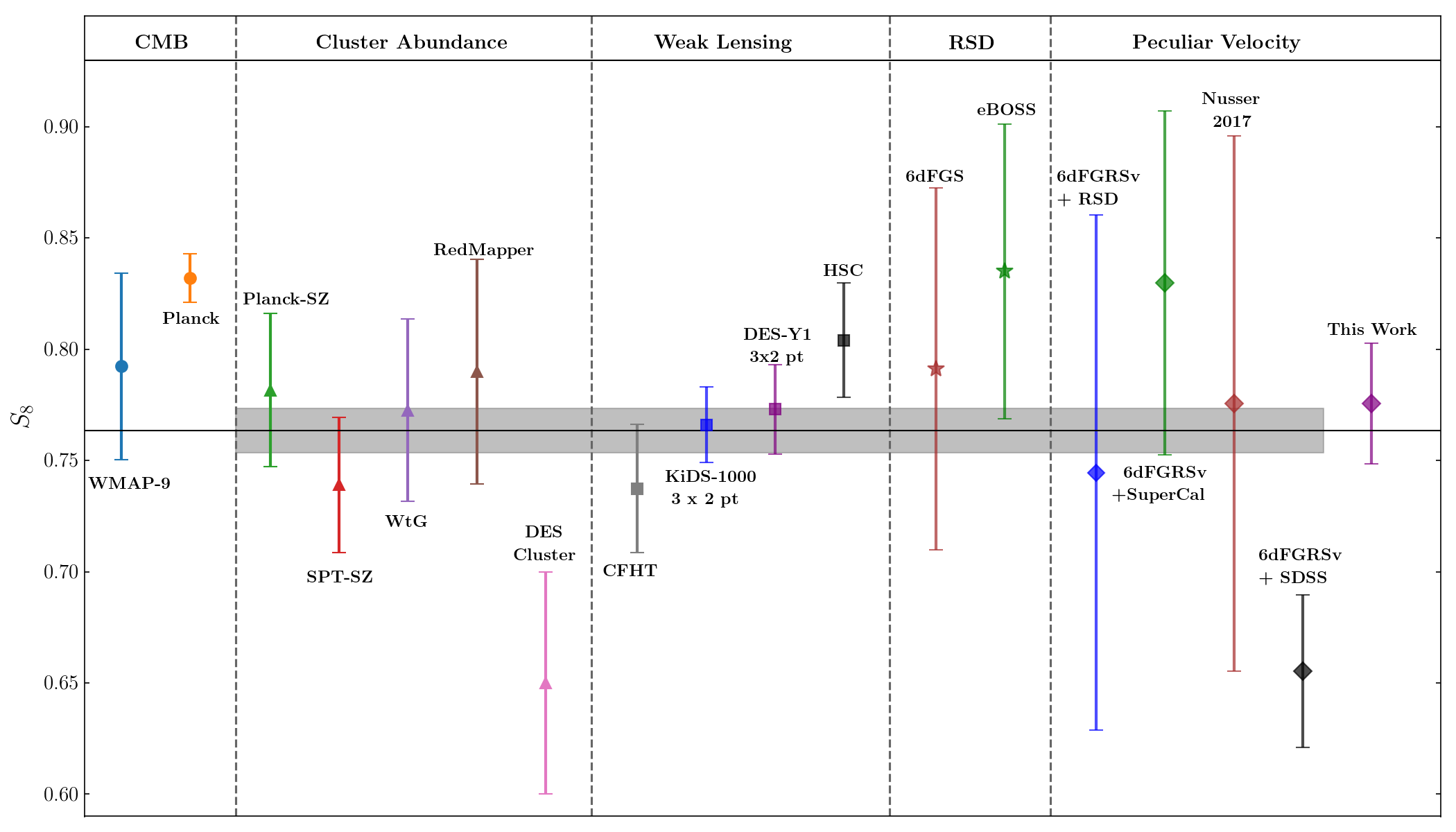}
    \caption{%
    Comparison of different literature results for $S_8 \equiv \sigma_8(z=0)(\Omega_m(z=0)/0.3)^{0.5}$. See Table \ref{tbl:fsigma8_comparison} and section \ref{ssec:compare_fsigma8} for details of these studies. The horizontal line corresponds to the uncertainty weighted mean for the measurement from the different datasets, excluding the CMB and our result. The shaded grey region is the weighted uncertainty for these same studies.
    } 
    \label{fig:compare_fsigma8}
\end{figure*}

\begin{table*}
  \centering
  \caption{Comparison of the $S_8$ results in the literature from various cosmological probes. Along with the value of $S_8 = \sigma_8 (\Omega_m/0.3)^{0.5}$, we also show the quantity and the value in terms of $\Omega_m$ and $\sigma_8$ that was reported in the original reference.}
  \begin{tabular}{l|l|c|c|c|l}
  \hline
  \hline
    Probe &  & Reported Quantity  & Reported Value & $S_8$ & Reference\\
    \hline
    \hline
   CMB & WMAP & MCMC Chains & --- & $0.792 \pm 0.053$ & \cite{wmap9} \\
   & Planck & MCMC Chains & --- & $0.832 \pm 0.013$ & \cite{planck_cmb_cosmo} \\
    \hline
    Cluster Abundance & Planck-SZ & $\sigma_8 (\Omega_m/0.31)^{0.3}$ & $0.774 \pm 0.034$ & $0.782 \pm 0.048$ & \cite{planck_sz} \\
    & SPT-SZ & $S_8$ & $0.739 \pm 0.041$ & $0.739 \pm 0.041$ & \cite{spt_sz} \\
    & WtG & $\sigma_8 (\Omega_m / 0.3)^{0.17}$ & $0.81 \pm 0.03$ & $0.773 \pm 0.053$ & \cite{wtg} \\
    & RedMapper & $S_8$ & $0.79^{+0.05}_{-0.04}$ & $0.79^{0.05}_{-0.04}$ & \cite{redmapper} \\
    & DES-Clusters & $S_8$ & $0.65\pm0.05$ & $0.65\pm0.05$ & \cite{des_clusters} \\
    \hline
    Weak Lensing & CFHT & $\sigma_8 (\Omega_m / 0.27)^{0.46}$ & $0.774^{+0.032}_{0.041}$ & $0.737 \pm 0.039$ & \cite{cfht} \\
    & KiDS-1000 & $S_8$ & $0.766^{+0.020}_{-0.014}$ & $0.766^{+0.020}_{-0.014}$ & \cite{KiDS1000_lensing} \\
    & DES-Y1 & $S_8$ & $0.773 \pm 0.026$ & $0.773 \pm 0.026$ & \cite{des_lensing} \\
    & HSC & $S_8$ & $0.804 \pm 0.032$ & $0.804 \pm 0.032$ & \cite{hsc_lensing} \\
    \hline
    RSD & 6dFGS & $f\sigma_8$ & $0.423 \pm 0.055$ & $0.791\pm0.103$ & \cite{6dF_rsd}\\
    & eBOSS & $\sigma_8$ & $0.838 \pm 0.059$ & 
    $0.835 \pm 0.066$ & \cite{eboss_rsd}\\
    \hline
    Peculiar Velocity & 6dFGRSv + RSD & $f\sigma_8$ & $0.384\pm0.054$(stat)$\pm0.061$(sys) & $0.707\pm 0.148$ & \cite{6df_cross_w_rsd}\\
    & 6dFGRSv + SuperCal & $f\sigma_8$ & $0.428^{+0.048}_{-0.045}$ & $0.780\pm 0.087$ & \cite{Huterer17}\\
    & Nusser 2017 & $f\sigma_8$ & $0.40\pm0.08$ & $0.776 \pm 0.120$ & \citet{Nusser_17}\\
    & 6dFGRSv + SDSS & $f\sigma_8$ & $0.338\pm0.027$ & $0.655 \pm 0.034$ & \citet{Said2020}\\
    & \textbf{This Work} & $\boldsymbol{f\sigma_8}$ & $\boldsymbol{0.400\pm0.017}$ & $\boldsymbol{0.776\pm 0.033}$ & This work\\
    \hline
  \end{tabular}
  \label{tbl:fsigma8_comparison}
\end{table*}

In this section, we compare our result to others from the literature. These include cosmological constraints obtained from other peculiar velocity analyses, from redshift space distortions, as well as from CMB anisotropies, cluster abundances and weak lensing. Table \ref{tbl:fsigma8_comparison} summarises the measurements from the literature and converts these to the parameter combination $S_8 \equiv \sigma_8 (\Omega_m/0.3)^{0.5}$ where appropriate assuming $\Omega_m = 0.3$. The comparison is shown in Figure \ref{fig:compare_fsigma8}.

We first compare our result $f \sigma_{8,\textrm{lin}} = \fsigmainv$ to the constraints on $f\sigma_8$ from other analyses of peculiar velocity in the local universe. In \cite{6df_cross_w_rsd}, the authors did a joint analysis of the peculiar velocity sample and the redshift space distortions in the galaxy redshifts of the 6dF galaxy survey to obtain, $f\sigma_8 = 0.384\pm0.054$(stat)$\pm0.061$(sys). In \citet{Huterer17}, the authors used the `SuperCal' sample of supernovae in addition to the 6dFGRSv catalogue to obtain the constraint, $f\sigma_8 = 0.428^{+0.048}_{-0.045}$. We also include the results from \citet{Nusser_17}, where the authors used velocity-density cross-correlation on the \textit{Cosmicflows-3} peculiar velocity data \citep{cosmicflows3} and the 2MASS redshift survey to obtain $f\sigma_8 = 0.40\pm0.08$. Finally, we include the results of \citet{Said2020}, who used velocity comparison with 2M++ reconstruction on a sample of Fundamental Plane peculiar velocity data from 6dF and SDSS to obtain, $f\sigma_8 = 0.338 \pm 0.027$. Note that this value is substantially different from our constraints, even though both works use the same 2M++ reconstruction field. The reason for this discrepancy is not yet clear. Also note that the all of the peculiar results compared here, except for this paper, are not independent of each other since they all include the same 6dFGRSv catalogue.

Different cosmological probes are sensitive to different combination of parameters. In particular, while peculiar velocities are sensitive to $f\sigma_8 = \Omega^{0.55}_m \sigma_8$, cosmological constraints from weak lensing are usually reported in terms of the parameter, $S_8 = \sigma_8 (\Omega_m/0.3)^{0.5}$. For easy comparison with other results, we convert the $f\sigma_8$ constraints to $f\sigma_8 / (0.3^{0.55}) = \sigma_8 (\Omega_m / 0.3)^{0.55} \approx S_8$. Converting our $f\sigma_8$ constraint in this manner, our result corresponds to $S_8 = 0.776 \pm 0.033$.

Redshift space distortions (RSD) are sensitive to the parameter combination $f\sigma_8$. We compared our results to the RSD results from the 6dFGRS \citep{6dF_rsd} and the results from the completed SDSS survey \citep{eboss_rsd}, including the eBOSS results. \citet{6dF_rsd} probed the redshift space distortions at low redshifts with an effective redshift of $z_{\textrm{eff}} = 0.067$. At that redshift, the value of $f\sigma_8 = 0.423 \pm 0.055$. For SDSS, the growth results are reported as $\sigma_8 = 0.838 \pm 0.059$. We converted this to a constraint on $S_8$ by assuming a prior, $\Omega_m = 0.298 \pm 0.022$ obtained from the Pantheon supernovae sample \citep{pantheon}. This results in a value of $S_8 = 0.835 \pm 0.066$, consistent with our results.

Comparing our results to the weak lensing constraints from DES-Y1 \citep{des_lensing}, KiDS-1000 \citep{KiDS1000_lensing} and the HSC \citep{hsc_lensing}, we find good agreement. 

Cluster abundances are another probe of cosmology that is sensitive to the cosmological parameter combination, $\sigma_8 \Omega^{\alpha}_m$, where, $\alpha$ is the local slope of the matter power spectrum \citep{cluster_cosmology}. Depending on the specific survey, $\alpha \sim 0.2-0.4$. We compare the results of 5 cluster abundance studies. Two of them Planck-SZ \citep{planck_sz} and SPT-SZ \citep{spt_sz} are based on Sunyaev-Zeldovich (SZ) clusters. SPT-SZ, the RedMapper \citep{redmapper} study and the DES cluster abundance \citep{des_clusters} give their results in terms of $S_8$. The Planck-SZ and Weighing the Giants (WtG) \citep{wtg} gives their results using $\alpha = 0.3$ and $\alpha = 0.17$ respectively. To convert these constraints into that on $S_8$, we use the $\Omega_m$ value inferred in these studies and add the uncertainties in quadrature.

Finally, we compare our results also to the results obtained from CMB anisotropies. We use the publicly available MCMC chains for Planck 2018 \citep{planck_cmb_cosmo} and Wilkinson Microwave Anisotropy Probe (WMAP) 9 year \citep{wmap9}  results to obtain the constraints on $S_8$. For the Planck results, we use the combination of TT,TE,EE+lowE+lensing results. 

Our result is in good agreement with a simple error weighted average of lower redshift results: the uncertainty-weighted value of $S_8$ of the measurements from the different datasets, excluding the CMB and our result, is $0.763\pm0.010$. (If we also include our result, we obtain $0.765\pm0.009$). While our result appears to be in tension with Planck, the difference is not statistically significant ($1.6\sigma$) Moreover, as discussed in section \ref{ssec:future_prospect}, the systematic uncertainties in our measurement have not been fully quantified at this time.

\subsubsection{Comparison of the bulk flow}\label{ssec:compare_bulk_flow}
\begin{figure}
    \centering
    \includegraphics[width=\linewidth]{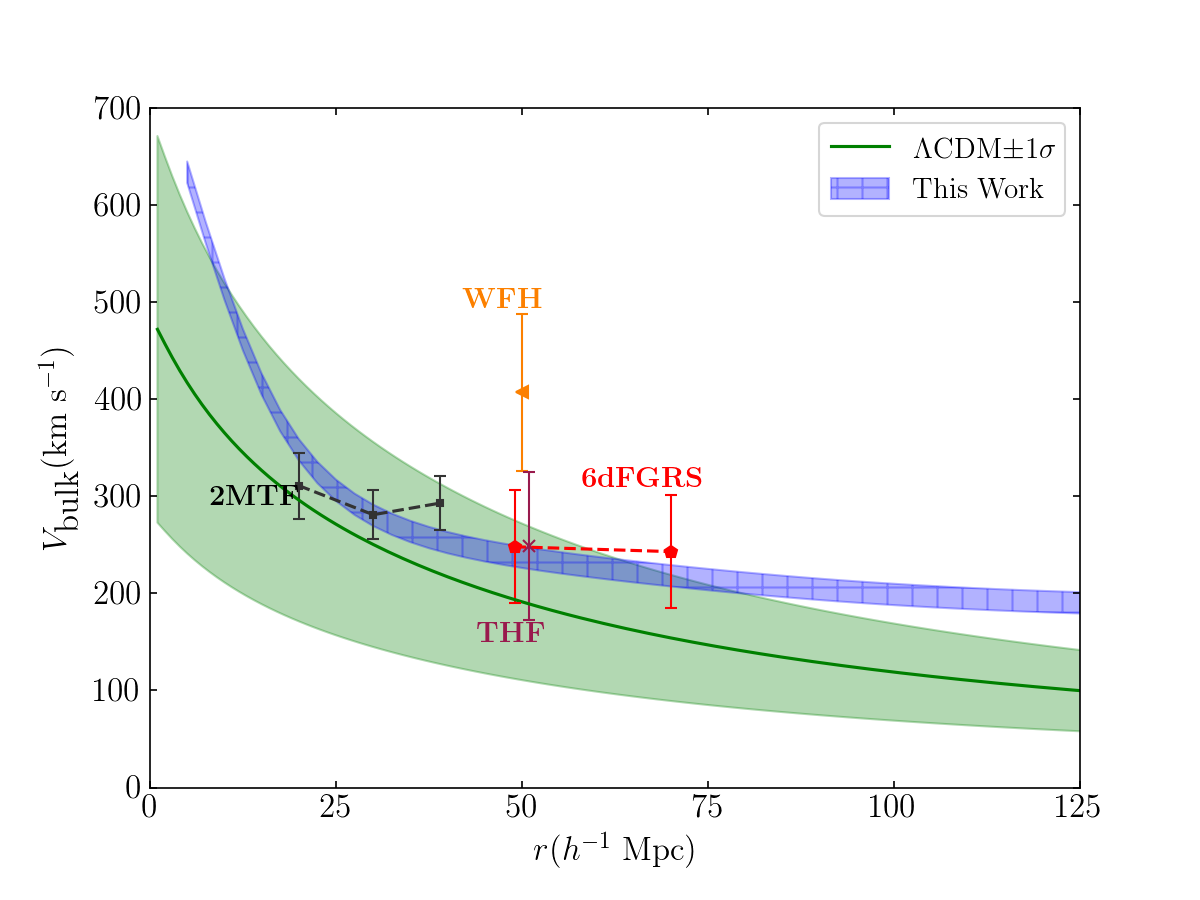}
    \caption{Comparison of the bulk flow amplitude. Our result for the bulk flow amplitude is compared to other results from the literature and to the $\Lambda$CDM prediction, which is calculated using a Gaussian filter of the given scale radius. The green shaded area shows the $68\%$ confidence region for $\Lambda$CDM predictions. Our bulk flow amplitude is calculated by adding the residual bulk flow inferred in the earlier sections to the Gaussian smoothed bulk velocity centered on the Local Group at different scales. Our result is shown with a hatched blue region.  The 2MTF \citep{2mtf_bulk_flow} bulk flow is denoted with a black symbols, 6dFGRS \citep{6dF_bulk_flow} with a red pentagon, THF \citep{turnbull_sn} with a brown cross, WFH \citep{wfh} with an orange triangle.}
    \label{fig:compare_bulk_flow_amplitude}
\end{figure}
\begin{table*}
  \centering
  \caption{Bulk flow results - comparison with other studies. We quote our bulk flow result at $40~ h^{-1}$ Mpc for easy comparison with other studies}
  \begin{tabular}{l|l|c|c|c|c|l}
  \hline
  \hline
    Work & Peculiar Velocity survey & Effective radius  & $|\mvec{V}_{\textrm{bulk}}|$ (km/s) & $l$ (deg) & $b$ (deg) & Reference\\
    \hline
    \hline
   6dFGRSv & 6dFGRSv & $40\ h^{-1}$ Mpc & $248 \pm 58$ & $318 \pm 20$ & $40 \pm 13$ & \cite{6dF_bulk_flow}\\
   2MTF & 2MTF& $40\ h^{-1}$ Mpc & $292 \pm 28$ & $296 \pm 16$ & $19 \pm 6$ & \cite{2mtf_bulk_flow}\\
   THF & A1 Supernovae& $50\ h^{-1}$ Mpc & $249 \pm 76$ & $319 \pm 18$ & $7 \pm 14$ & \cite{turnbull_sn}\\
   WFH & COMPOSITE & $40\ h^{-1}$ Mpc & $407 \pm 81$ & $287 \pm 9$ & $8 \pm 6$ & \cite{wfh}\\
   \textbf{This Work} & \textbf{A2 Supernovae + SFI++} & $\boldsymbol{40\ h^{-1}}$ \textbf{Mpc} & $\boldsymbol{\vbulkinv}$ & $\boldsymbol{293 \pm 5}$ & $\boldsymbol{14 \pm 5}$ & ---- \\
    \hline 
  \end{tabular}
  \label{tbl:bulk_flow_comparison}
\end{table*}
\begin{figure}
    \centering
    \includegraphics[width=\linewidth]{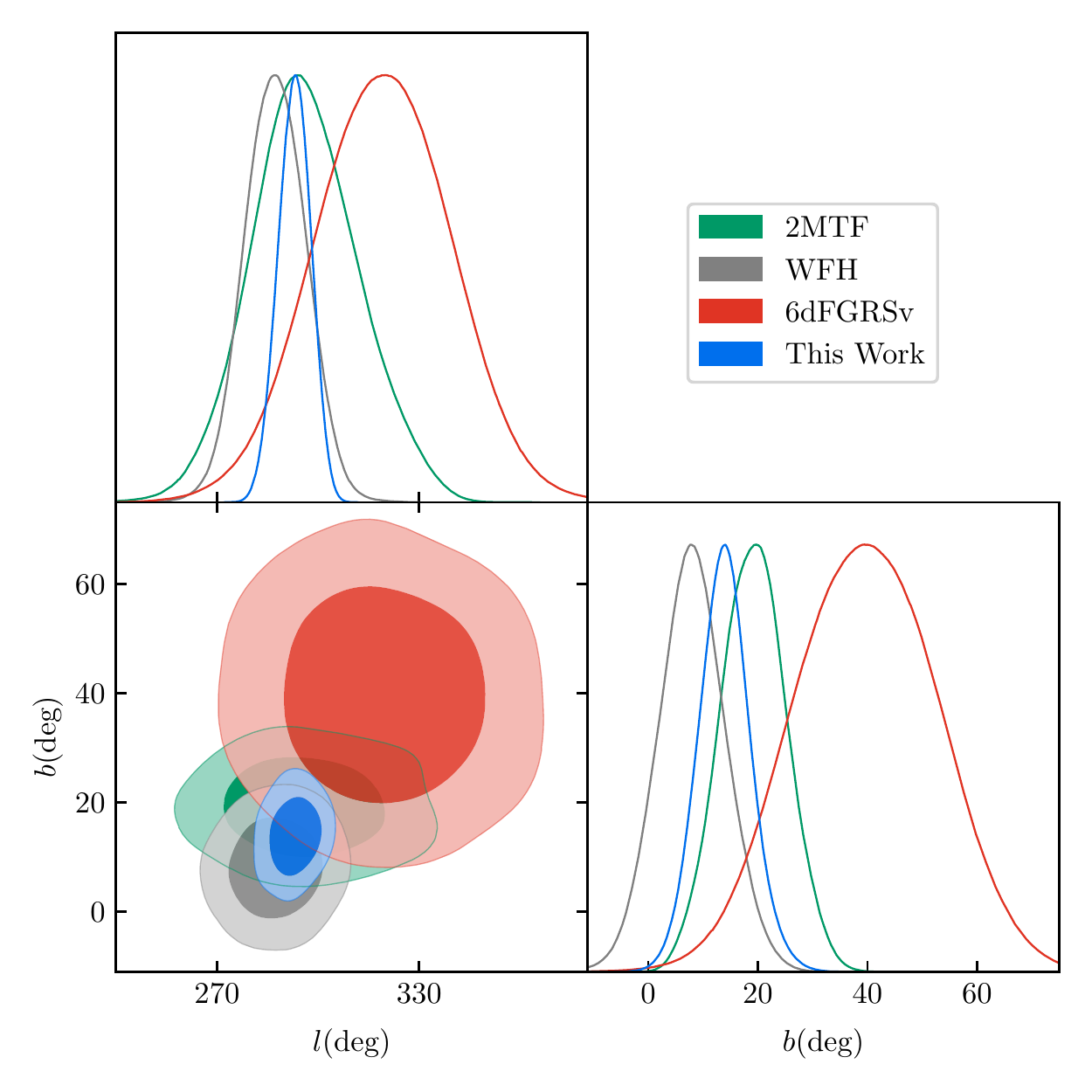}
    \caption{Comparison of the direction of the bulk flow at a depth of $\sim 40\ h^{-1}$ Mpc. We obtain a bulk flow in the direction $(l,b) = (293^{\circ} \pm 5^{\circ}, 14^{\circ} \pm 5^{\circ})$ for our reconstruction. For the purposes of illustration, the other works (2MTF, WFH and 6dFGRSv) are plotted as a normal distribution in $l, b$ with the quoted uncertainties.  The dark and light shaded regions correspond to the $68\%$ and $95\%$ confidence intervals respectively}
    \label{fig:bulk_flow_direction}
\end{figure}
The bulk flow in the local universe has been studied in the literature by many groups \citep[see e.g.][]{Carrick_et_al, 6dF_bulk_flow, 2mtf_bulk_flow}. In this section, we compare our results to the predictions from the $\Lambda$CDM model and to other results in the literature.

One can use linear perturbation theory to calculate the expected bulk flow in a $\Lambda$CDM universe. The variance of the bulk flow on a scale, $R$, is given as \citep{bulk_flow_expectation}, 
\begin{equation}
    \sigma^{2}_B(R) = \frac{H^2_0 f^2}{2\pi^2} \int_0^{\infty} dk P(k)\widetilde{W}^2(k,R)\;,
\end{equation}
where $P(k)$ is the matter power spectrum and $\widetilde{W}$ is the window function used to smooth the field at the scale, $R$. We calculate the matter power spectrum using the publicly available CAMB software \citep{camb}. 

The distribution of velocity on a scale, $R$, with standard deviation, $\sigma_B$ is given by the Maxwellian distribution if the density field is Gaussian. On large-scales, where linear theory holds, this is a valid assumption. Hence, the distribution of bulk flow velocity, $V$, for $\Lambda$CDM universe is given as,
\begin{equation}
    P(V)\,\mathrm{d}V = \sqrt{\frac{2}{\pi}}\bigg(\frac{3}{\sigma^2_B}\bigg)^{3/2}V^2\exp\bigg(-\frac{3 V^2}{2\sigma^2_B}\bigg)\,\mathrm{d}V\;.
\end{equation}
We plot the mean and standard deviation of this distribution as a function of the scale, $R$, in Figure \ref{fig:compare_bulk_flow_amplitude}. These results are calculated assuming $\Omega_m = 0.3$. We compare our results along with other results of bulk flow in the literature. At a radius of $125\ h^{-1}$ Mpc, up to which 2M++ has high all-sky completeness, the mean and standard deviation of the predicted bulk flow for $\Lambda$CDM is $100$~km~s$^{-1}$ and $42$~km~s$^{-1}$ respectively. 

At large radius, the predicted bulk flow from our reconstruction at $125~h^{-1}$ Mpc, $V_{\text{bulk}}(125h^{-1}$ Mpc) $=189 \pm 11$ km/s, seems to be slightly higher than the linear theory predictions, but the $\sim 2.1 \sigma$ difference is formally only marginally significant. This bulk flow consists of a $\sim 170$ km/s external bulk flow and a $19$ km/s internal bulk flow from the reconstruction. Note that the bulk flows in spheres of increasing radius are highly correlated. Also recall that our model of the contributions to the flow field due to external sources is simplified to the lowest order, a dipole, which is fit to peculiar velocity data that are clustered close to the origin and which do not fill the volume. For example, it is possible that a massive attractor, hidden behind the Galactic Plane and so not included in our 2M++ model, but at a distance less than $125~h^{-1}$ Mpc, could account for some of residual bulk motion. If that were the case, then the extrapolation of the dipole $V_{\text{ext}}$ model to $125~h^{-1}$ Mpc would not be correct. Therefore, at this time, we do not assign a high significance to this discrepancy.

At $r \approx 0$, our estimates are consistent with the Local Group velocity. The velocity of the Local Group as inferred in our reconstruction is $706$ km/s in the direction, $l = 277^{\circ}$ and $b = 35^{\circ}$ in agreement with the CMB dipole, given the expected scatter of 150 km/s when comparing halo peculiar velocities with the predictions of linear theory. Note that the result for the predicted bulk flow variance is also not sensitive to changes in the value adopted for $\Omega_m$. A change of $5\%$ changes the bulk flow by $<0.5\%$. In summary, the results for the external bulk flow obtained from our reconstruction are not inconsistent with the $\Lambda$CDM predictions. We also compare our results with other studies of bulk flow in the literature. The details of these other studies are given in Table \ref{tbl:bulk_flow_comparison}. In this comparison, we quote our bulk flow results at $R = 40\ h^{-1}$ Mpc. \citet{Said2020} used a Fundamental Plane peculiar velocity sample to infer an external bulk flow with the 2M++ reconstruction. They inferred a bulk flow of $168 \pm 11$ km s$^{-1}$ in the direction $l=304^{\circ} \pm 4^{\circ}$, $b=1^{\circ} \pm 4^{\circ}$, which is in remarkable agreement with our results. The direction of the bulk flow as found in this study is also similar to what has been found before in other studies. We compare some of these in Figure \ref{fig:bulk_flow_direction}. We do not include the results from \citet{Said2020} in Figures \ref{fig:compare_bulk_flow_amplitude} and \ref{fig:bulk_flow_direction} since they are virtually indistinguishable from our results. 

\subsection{Future Prospects}\label{ssec:future_prospect}

In the near future, we anticipate an order-of-magnitude increase in peculiar velocity data from new surveys. The ``Transforming Astronomical Imaging surveys through Polychromatic Analysis of Nebulae'' (TAIPAN) survey \citep{taipan} will acquire the distances to $\sim 45,000$ galaxies up to $z \sim 0.1$ in the southern sky using the FP relation. The Widefield ASKAP L-band Legacy All-sky Blind Survey \citep[WALLABY,][]{wallaby} survey is a HI-survey which will observe 3 quarters of the sky. Using the Tully-Fisher relation, it is expected to acquire distances to $\sim 40,000$ galaxies \citep{pec_vel_forecast}. In comparison, at present, the largest Tully-Fisher catalogue is the SFI++ catalogue with $\sim 4,500$ galaxies. It has been forecast that using a combination of the WALLABY and the TAIPAN peculiar velocity data, the constraints on $f \sigma_8$ will reach $\sim 3\%$ \citep{pec_vel_forecast, pec_vel_forecast2}. There will also be an increase in the number of low-redshift Type Ia supernovae usable for peculiar velocity studies. The full Foundation supernovae sample will consist of up to $800$ supernovae at $z \lesssim 0.1$ \citep{foundation1, foundation2}. In the near future, Large Synoptic Survey Telescope (LSST) will also start taking data. It is expected to greatly increase the number of supernovae known in the local universe, although many will be at redshifts $z \gtrsim 0.2$ and will therefore have large uncertainties \citep{lsst_pec_vel}. Together, the use of these peculiar velocity estimates could provide us with unprecedented constraints on the growth rate in the local universe.

Given the statistical precision of peculiar velocity studies, it would be timely to more clearly understand the systematics of the density-velocity comparison. \citet{Carrick_et_al} used N-body simulations to show that, when dark matter haloes are used as tracers of the density field, the inverse reconstruction procedure used here should have biases of order of 1\%. However, in practice, a few approximations were made during the reconstruction. For example, luminosity-weighting is used as a proxy for halo mass, where every galaxy is assigned a weight proportional to their absolute $K$-band luminosity. The galaxy bias is assumed to be linear on a scale where it is known to be non-linear, with the bias factor fit from the data. The systematic errors introduced due to these approximations have not been quantified.

Improvement in the methods of analysis may also tighten these constraints. Forward-modelled reconstruction is a promising framework for the analysis of the large-scale structure. In \citet{virbius}, a forward-modelled approach, {\sc virbius}, was introduced to analyse the 3-dimensional velocity field by jointly inferring the distances to the peculiar velocity data. {\sc virbius} was used in \citet{virbius_cosmicflows} to analyse the CosmicFlows-3 data.  It would also be interesting to compare the velocity field of the forward modelled reconstruction scheme, \borg{} \citep{BORG_original, BORG_PM} to study the peculiar velocity field of the local universe. Non-linear structure formation models such as Second order Lagrangian Perturbation Theory \citep[2LPT,][]{2lpt}, Particle-Mesh \citep[See e.g., ][]{hockney_eastwood} and COmoving Lagrangian Acceleration \citep[COLA,][]{cola} can be incorporated into \borg{}, likely providing a better approximation to the non-linear velocity field. 
\section{Summary}\label{sec:summary}

In this work, we used peculiar velocity analysis to infer the cosmological parameter combination $f\sigma_8$ and the bulk flow in the local universe. We compiled a new peculiar velocity catalogue of low-$z$ Type Ia supernovae, called the Second Amendment (A2) sample. We also used the SFI++ and the 2MTF Tully-Fisher catalogues for our analysis. We used an inverse reconstruction scheme used in \citet{Carrick_et_al} to compare the predicted velocities from the reconstruction to the observations in order to infer $f\sigma_8$ and the bulk flow. To make this comparison, we introduced a variant of the original forward likelihood method, in which the distances to the peculiar velocity tracers are fitted jointly with the flow model and hence, do not require prior calibration. The comparison yielded $f\sigma_{8, \textrm{lin}} = \fsigmainv$, with $\sim ~ 4\%$ statistical uncertainties on the value of $f\sigma_8$. These results are consistent with other low redshift results from the literature, as shown in section \ref{ssec:compare_fsigma8}. We also fit for an external bulk flow which is not accounted for in our reconstruction process. We compare our constraint of the bulk flow with the $\Lambda$CDM prediction in Figure \ref{fig:compare_bulk_flow_amplitude}. With our reconstruction method, we obtain a residual bulk flow of $\vextinv{}$ km s$^{-1}$ in the direction $l = \lextinv{}, b=\bextinv$. At an effective radius of $40 h^{-1}$ Mpc, this corresponds to a bulk flow of $\vbulkinv$ km s$^{-1}$ in the direction $l = \lbulkinv, b=\bbulkinv$ for our reconstruction scheme. 

\section*{Acknowledgements}

This work has been done as part of the activities of the Domaine d'Int\'{e}r\^{e}t Majeur (DIM) ``Astrophysique et Conditions d'Apparition de la Vie'' (ACAV), and received financial support from R\'{e}gion Ile-de-France. We acknowledge support from a Mitacs Globalink Award. GL acknowledges financial support from the ANR BIG4, under reference ANR-16-CE23-0002.  MH acknowledges the support of an NSERC Discovery Grant.
This  work  was granted  access  to  the  HPC  resources  of  CINES  (Centre  Informatique National de l'Enseignement Sup\'erieur) under the allocation  A0070410153  made by GENCI and  has  made  use  of  the  Horizon  cluster  hosted  by  the  Institut  d'Astrophysique  de  Paris. We thank St\'{e}phane Rouberol for smoothly running the Horizon cluster. GL thanks the University of Helsinki for hospitality. We used the \texttt{getdist} package \citep{getdist} for plotting the MCMC samples. This work is done within the Aquila Consortium\footnote{\url{https://www.aquila-consortium.org/}}. This research was enabled in part by the computing support provided by Compute Ontario\footnote{\url{https://computeontario.ca/}} and Compute Canada\footnote{\url{https://www.computecanada.ca/}}.

\section*{Data availability statement}
The Second Amendment (A2) supernovae catalogue that was compiled for this work will be made available as online supplementary material upon acceptance. The SFI++ and the 2MTF catalogues are publicly available with their respective publications. The 2M++ reconstruction used in this work is publicly available at \url{https://cosmicflows.iap.fr/}. 



\bibliographystyle{mnras}
\bibliography{sn_peculiar_velocity} 

\appendix

\section{Testing the modified forward likelihood with simulated supernovae data}\label{sec:sn_mock}

\begin{figure}
    \centering
    \includegraphics[width=\linewidth]{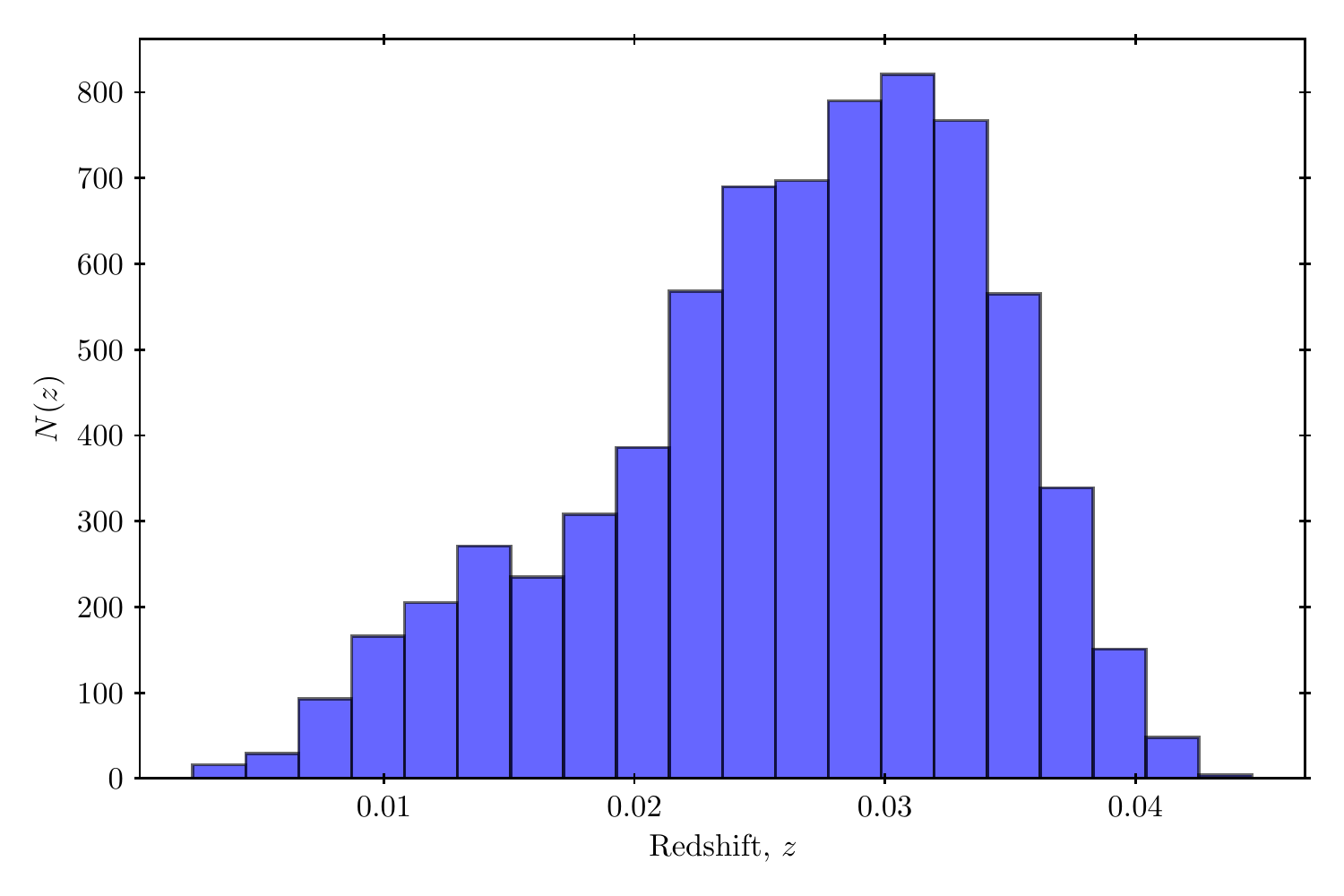}
    \caption{The redshift distribution of the simulated supernovae after the apparent magnitude cut of $16.5$.}
    \label{fig:sn_sim_z_dist}
\end{figure}

\begin{figure*}
    \centering
    \includegraphics[width=\linewidth]{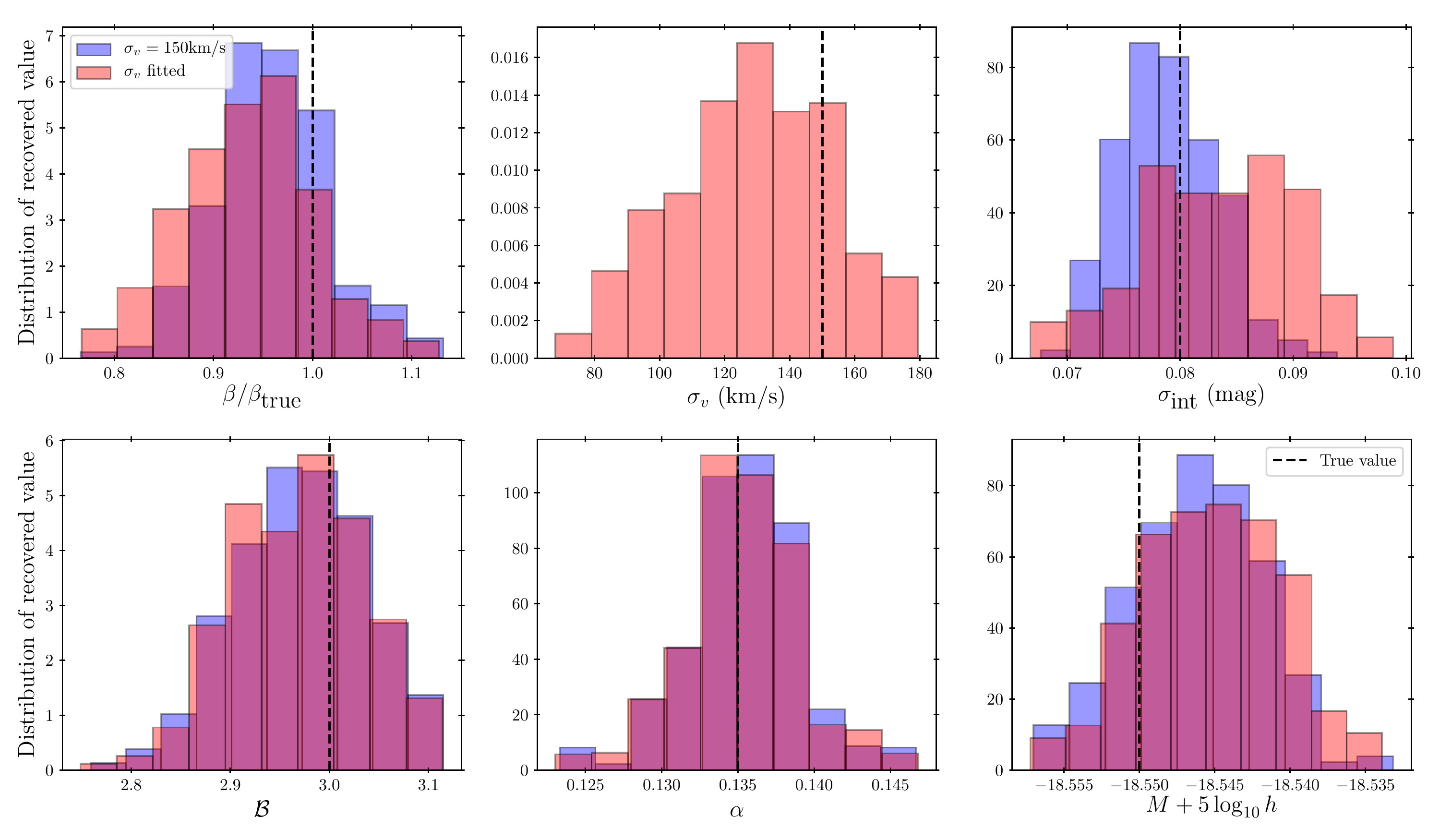}
    \caption{The recovered value of the various parameters from the simulated supernovae sample. We plot the histogram of the recovered values for the various parameters weighted by its uncertainty. The black dashed vertical line shows the true value of the parameter. For the adopted smoothing of $4~h^{-1}$ Mpc, we expect $\beta/\beta_{\textrm{true}} = 0.95$, see text for further details.}
    \label{fig:sn_mock}
\end{figure*}

\begin{figure}
    \centering
    \includegraphics[width=\linewidth]{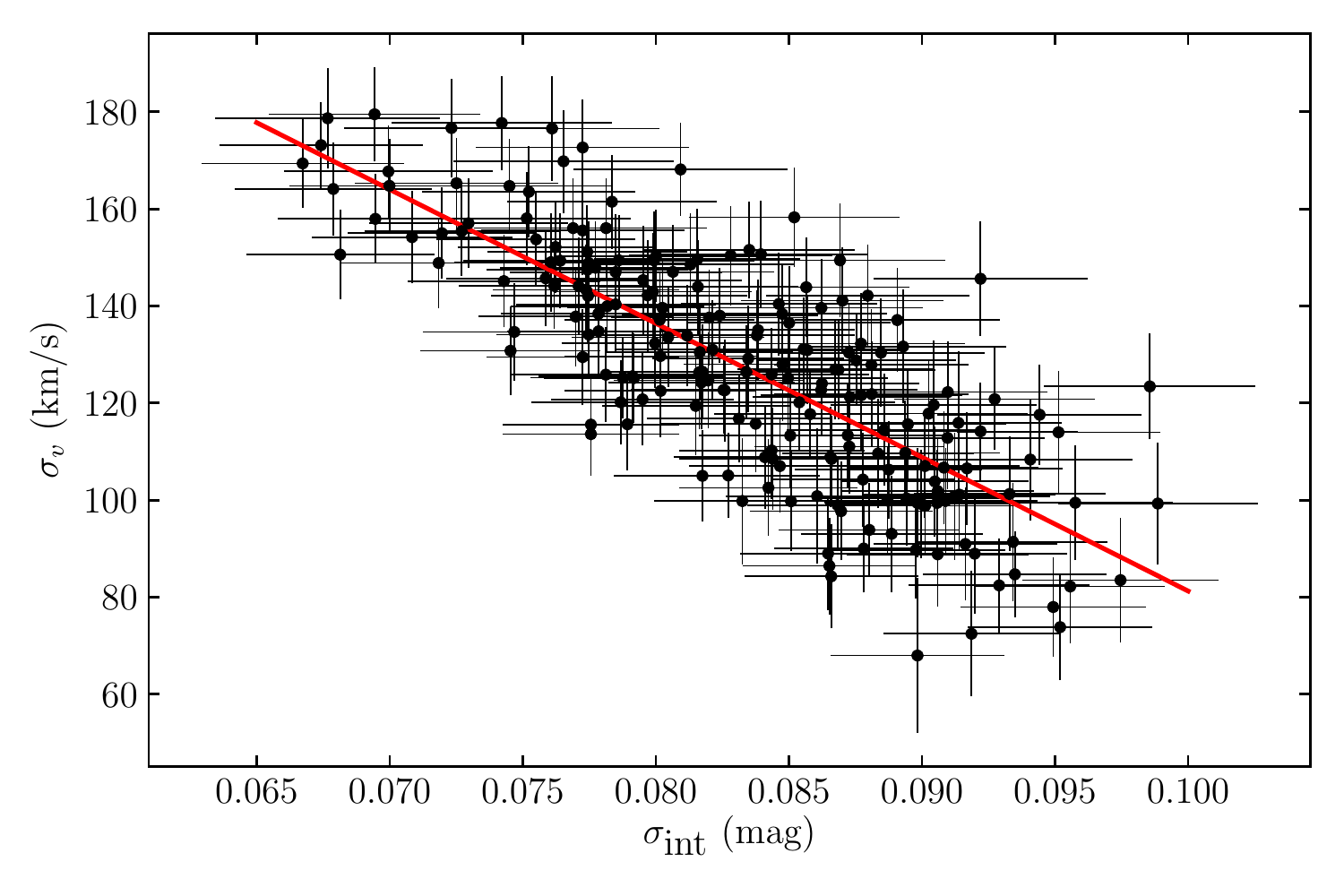}
    
    \caption{Degeneracy in the recovered value of $\sigma_v$ and $\sigma_{\text{int}}$. We plot a rough best fit line, $\sigma_v = [137 + 220(\sigma_{\text{int}}/0.08 - 1)]$ km/s to indicate the direction of the degeneracy. Simultaneously fitting $\sigma_v$ and $\sigma_{\text{int}}$ can result in significant bias in the recovered value of these parameters.}
    \label{fig:sigma_int_sigv_degeneracy}
\end{figure}

In this work, we used a modified forward likelihood method to jointly infer both the flow model and the distances to the supernovae. In this Appendix, we use this method on simulated  data, where we know the true values of the global parameters of the SALT2 fitting formula.

To create the simulated data, we used a simulation from the VELMASS N-body simulation \footnote{For more details on the simulation we used, see \citet{halo_painting, borg_mdn}}. The simulation we use was performed in a cubic box of size $2~h^{-1}$ Gpc with a total of $2048^3$ particles with mass $9.387\times 10^{10}~h^{-1}M_{\odot}$. The cosmological parameters used are as follows: $\Omega_m= 0.315, \Omega_b = 0.049, H_0 = 68$ km s$^{-1}$ Mpc$^{-1}$, $\sigma_8 = 0.81$, $n_s = 0.97$ and $Y_{\text{He}} = 0.248$. We identified the halos in the simulation with the ROCKSTAR halo finding software \citep{rockstar}. We then populate the dark matter halos with simulated supernovae. To each supernova, we assign a value of the stretch ($x_1$) and color ($c$) parameter according to a distribution which describes the data from the Foundation supernovae survey well. We then use values of the global parameters as follows, $M = -18.55, \alpha=0.135, \mathcal{B} = 3.00, \sigma_{\textrm{int}} = 0.08$ mag. Using the true distance of the halo (and hence the true distance modulus, $\mu$) and adding a Gaussian noise with a scatter of $\sigma_{\textrm{int}}$, we can then measure the value of $m_B$ using equation \eqref{eqn:tripp}. We then apply an apparent magnitude cut of $m_B < 16.5$. The redshift distribution of the resulting simulated supernovae sample is shown in Figure \ref{fig:sn_sim_z_dist}. From this data, we create $200$ realizations of $500$ randomly selected mock supernovae.

The density field of the VELMASS simulation is calculated using the Cloud-in-cell algorithm with grids of size $1.953~h^{-1}$ Mpc. We then smooth this density field with a Gaussian filter of a given size and estimate the velocity field from the smoothed density using the linear perturbation theory prediction, equation \eqref{eqn:linear}. When the Gaussian smoothing is done in conjunction with the CIC gridding, the effective smoothing length is different from the base Gaussian smoothing length. The CIC kernel in Fourier space is given as \citep{hockney_eastwood, cic_kernel}, 
\begin{equation}
    W_{\text{CIC}}(\mvec{k}; l) = \prod_{i\in\{x,y,z\}}\bigg[\frac{\sin^2(k_i l/2)}{(k_i l/2)^2}\bigg],
\end{equation}
where $l$ is the grid spacing. Taylor expanding the CIC kernel, we get
\begin{equation}
    \ln[W_{\text{CIC}}(\mvec{k};l)] \approx -\frac{k^2}{2}\bigg(\frac{l^2}{6}\bigg) + \mathcal{O}(k^4l^4).
\end{equation}
Thus the CIC kernel has an effective Gaussian smoothing scale of $l/\sqrt{6}$. For a density field calculated with nearest grid-point (NGP) and Triangular Shaped
Cloud (TSC) algorithms, the equivalent Gaussian smoothing length calculated in the same way turns out to be $l/\sqrt{12}$ and $l/2$ respectively. Therefore, when smoothing the CIC gridded density field with a Gaussian filter of size, $R$, the effective Gaussian smoothing scale is given as, $R^2_{\text{eff}} = R^2 + l^2/6$. 

For fitting the parameters with the forward likelihood code, we use the density field smoothed at $R_{\text{eff}} = 4~h^{-1}$ Mpc. The velocity field is calculated using linear perturbation theory prediction, equation \eqref{eqn:linear}. We jointly infer the global parameters of the Tripp parameterization and the flow model using the method of Section \ref{sssec:fwd_lkl_joint} for the $200$ realizations of the mock supernovae. Note that all these realizations are not completely independent of each other. There are a total of $\sim 9000$ mock supernovae after applying the apparent magnitude cut. Therefore, there must be repeated sampling of supernovae in our realizations. The results of the recovered value of the forward likelihood fit is shown in Figure \ref{fig:sn_mock}. We plot the distribution of the recovered mean weighted by their uncertainty for the different parameters. We fit the parameters in two ways - {\it i)} we fix $\sigma_v = 150$ km/s, {\it ii)} we fit $\sigma_v$ as a free parameter. 

As can be seen from Figure \ref{fig:sn_mock}, we recover unbiased value of the SNe parameters, $\alpha$ and $\mathcal{B}$ using both the methods. When simultaneously fitting $\sigma_{v}$, we see that the recovered value of $\sigma_{\text{int}}$ has a significant scatter. This results from the fact that jointly fitting $\sigma_v$ and $\sigma_{\text{int}}$ results in a significant degeneracy between these parameters. This is shown in Figure \ref{fig:sigma_int_sigv_degeneracy} where we plot the recovered value of $\sigma_v$ and $\sigma_{\text{int}}$ for the $200$ realizations. Therefore, if one simultaneously fits $\sigma_v$ and $\sigma_{\text{int}}$, it can result in significantly biased results. For the parameter, $\beta / \beta_{\text{true}}$, we recover a weighted mean of $0.96$ with a mean uncertainty of $0.04$, which agrees with the scatter from simulation to simulation around the weighted mean. As shown in \citet{Carrick_et_al}, the value of $\beta$ obtained by using the smoothed density field from haloes is not the same as the value of $\beta$ obtained, at the same smoothing length, using the reconstruction procedure in which the haloes are iteratively moved from theory positions in redshift space to their reconstructed positions in configuration space. In fact, using the haloes at their true locations results in a value of $\beta$ that is biased low by $\sim 5\%$ at $4~h^{-1}$ Mpc, as we also find here. Since the comparison in this paper uses the \citet{Carrick_et_al} density field after reconstruction, we expect this to be unbiased. Finally, there is no significant bias in the value of $M$ obtained using the two methods. The weighted value for all the simulations is $-18.546 \pm 0.005$ mag using $\sigma_v = 150$ km/s and $-18.545 \pm 0.005$ mag when we simultaneously fit $\sigma_v$.



\bsp	
\label{lastpage}
\end{document}